\documentclass[aps,prb,twocolumn,showpacs,amsmath,amssymb]{revtex4-1}

\usepackage{amsmath}
\usepackage{amssymb}
\usepackage[colorlinks,plainpages=false,linkcolor=black,urlcolor=blue,citecolor=blue,pdfpagemode=UseNone,pdfstartview=FitH]{hyperref}
\usepackage[T1]{fontenc}
\usepackage[utf8]{inputenc}
\usepackage[english]{babel}
\usepackage[left=2.5cm,right=2.5cm,top=2cm,bottom=3.0cm,a4paper]{geometry}
\usepackage{color}
\usepackage{float}

\usepackage{graphicx}
\graphicspath{{Figs/}}
\usepackage{mathtools}

\newcommand{\mi}{\rm i}

\newcommand{\cuuuu}{\chi^{\uparrow\uparrow}_{\uparrow\uparrow}}
\newcommand{\cdddd}{\chi^{\downarrow\downarrow}_{\downarrow\downarrow}}
\newcommand{\cuudd}{\chi^{\uparrow\uparrow}_{\downarrow\downarrow}}
\newcommand{\cdduu}{\chi^{\downarrow\downarrow}_{\uparrow\uparrow}}
\newcommand{\cuddu}{\chi^{\uparrow\downarrow}_{\downarrow\uparrow}}
\newcommand{\cduud}{\chi^{\downarrow\uparrow}_{\uparrow\downarrow}}
\newcommand{\cudud}{\chi^{\uparrow\downarrow}_{\uparrow\downarrow}}
\newcommand{\cdudu}{\chi^{\downarrow\uparrow}_{\downarrow\uparrow}}
\newcommand{\csigma}{\chi^{\sigma_1\sigma_2}_{\sigma_3\sigma_4}}

\newcommand{\Uuuuu}{U^{\uparrow\uparrow}_{\uparrow\uparrow}}
\newcommand{\Udddd}{U^{\downarrow\downarrow}_{\downarrow\downarrow}}
\newcommand{\Uuudd}{U^{\uparrow\uparrow}_{\downarrow\downarrow}}
\newcommand{\Udduu}{U^{\uparrow\uparrow}_{\downarrow\downarrow}}

\newcommand{\Uudud}{U^{\uparrow\downarrow}_{\uparrow\downarrow}}
\newcommand{\Ududu}{U^{\downarrow\uparrow}_{\downarrow\uparrow}}

\newcommand{\diagram}[1]{\vcenter{\hbox{\includegraphics[scale=0.5]{#1}}}}
\newcommand{\diagramdef}[1]{\vcenter{\hbox{\includegraphics[scale=0.275]{#1}}}}

\begin{document}

\title{Anisotropic spin fluctuations in Sr$_2$RuO$_4$: role of spin-orbit coupling and induced strain}

\author	{Sergio Cobo$^{1}$}
\author	{Felix Ahn$^{2}$}
\author	{Ilya Eremin$^{2}$}
\author	{Alireza Akbari$^{1,3,4}$}\email{alireza@apctp.org}

\affiliation	{$^{1}$Asia Pacific Center for Theoretical Physics (APCTP), Pohang, Gyeongbuk, 790-784, Korea}
\affiliation	{$^2$Institut f\"ur Theoretische Physik III, Ruhr-Universit\"at Bochum, D-44801 Bochum, Germany}
\affiliation	{$^{3}$Department of Physics, POSTECH, Pohang, Gyeongbuk 790-784, Korea}
\affiliation	{$^{4}$Max Planck POSTECH Center for Complex Phase Materials, POSTECH, Pohang 790-784, Korea} 

\begin{abstract}
We analyze the spin anisotropy of the magnetic susceptibility of Sr$_2$RuO$4$ in presence of spin-orbit coupling and anisotropic strain using quasi-two-dimensional tight-binding parametrization fitted to the ARPES results.
Similar to the previous observations we find the in-plane polarization of the low ${\bf q}$ magnetic fluctuations and the out-of-plane polarization of the incommensurate magnetic fluctuation at the nesting wave vector ${\bf Q}_1 = (2/3 \pi ,2/3 \pi)$ but also nearly isotropic fluctuations near ${\bf Q}_2=(\pi/6,\pi/6)$. Furthermore, one finds that apart from the high-symmetry direction of the tetragonal Brillouin zone the magnetic anisotropy is maximal, i.e. $\chi^{xx} \neq \chi^{yy} \neq \chi^{zz}$. This is the consequence of the orbital anisotropy of the $xz$ and $yz$ orbitals in the momentum space. We also study how the magnetic anisotropy evolves in the presence of the strain and find strong Ising-like ferromagnetic fluctuations near the Lifshitz transition for the $xy$-band.
\end{abstract}
\date{\today}
\maketitle
%
%
{\it Introduction:}
Since its discovery in 1994, Strontium Ruthenate, Sr$_2$RuO$_4$, 
has been one of the few widely studied triplet superconductors\cite{Maeno,Mackenzie}. 
Many experimental results provide indirect evidence for a triplet state with a broken time-reversal symmetry and odd-parity Cooper pairs, although the `smoking gun' experiment is still missing. Among these are the Knight shift measurements\cite{Ishida,Ishida2} that are in agreement with polarized neutron scattering experiments\cite{Duffy}. There are also indications of the broken time-reversal symmetry by polar Kerr~Effect measurements\cite{Xia}. Further studies have been performed to describe the unconventional superconducting state in Sr$_2$RuO$_4$\cite{Maeno2,Mackenzie,Kidwingira,Nelson,Thomale,Wei,Rastovski,Maeno3,Akbari,Annett}. Regarding the microscopic mechanism of the Cooper-pairing it is believed to be driven by the spin and charge fluctuations~\cite{Eremin,Raghu:2010aa,Tsuchiizu2015} where the multiorbital character of the bands plays an important role. 
Recent experiments  reveal that the transition temperature 
to the superconducting state in Sr$_2$RuO$_4$ can be enhanced locally if pressure is applied\cite{Taniguchi,Ying,Hicks}. 
A local enhancement of the transition at 
 $\sim 1$K 
was observed near lattice deformations\cite{Ying} and  
more specifically,  recent developments indicate an  enhancement of  $T_c$ up to  \mbox{$T_c=3.4$K} under the application of pressure  in the direction of the \mbox{$a$-axis}\cite{Steppke}. 
In addition, a phase transition from the superconducting state to a spin density wave state was later  predicted for even larger values of strain\cite{Liu}. 
This is a remarkable result, since Sr$_2$RuO$_4$ is generally known to be sensitive to disorder\cite{Haselwimmer}.

One of the intriguing complications of Sr$_2$RuO$_4$ is its multiorbital and multiband character  as the Fermi surface (FS) of this system shows three bands and very likely not all of the FS pockets are contributing equally to the Cooper-pairing\cite{Agterberg,Scaffidi}. For example, it was argued that the two mostly quasi-one-dimensional bands ($xz$ and $yz$ bands) with incommensurate AF spin and charge fluctuations may be driving superconductivity\cite{Raghu:2010aa,Firmo:2013aa}. At the same time, other groups argue in favor of the  dominant contribution to the Cooper-pairing from the large electron pocket of the $xy$-character. It is centered near the $\Gamma$-point of the Brillouin Zone (BZ)\cite{Liu} ($\gamma$-band) and lie close  for the Van~Hove~singularity near $(\pi,0)$ and $(0,\pi)$ points of the BZ\cite{Nomura:2000aa,Wang:2013aa,Yanase:2003aa,Nomura:2002aa}. Furthermore, the role of orbital versus band description of superconductivity was also discussed\cite{Mineev}. 
The $\gamma$-band is believed to be mainly affected by the application of anisotropic strain, consequently, the increase of $T_c$ upon strain is  mainly attributed to this band\cite{Thomale}. Further complexity in Sr$_2$RuO$_4$ comes from the relatively strong spin-orbit coupling in this system as confirmed by NMR\cite{Ishida2}, neutron scattering\cite{Sidis} and spin-resolved ARPES\cite{Veenstra2014} experiments. Furthermore, spin-orbit coupling plays also an important role in determining the characteristics of  the superconducting state\cite{Ng,Eremin2,Annett}.
\\

 In this paper, we study the evolution of the magnetic anisotropy of the spin susceptibility in Sr$_2$RuO$_4$ in the presence of spin-orbit coupling  and anisotropic strain using the tight-binding model fitted to the available ARPES results\cite{Zabolotnyy}. We compute the components of the spin susceptibility to obtain the full structure of the spin anisotropy within the itinerant description for the Hubbard-Hund type of the interaction model. Our results show clear anisotropy of the different components of the spin susceptibility enhanced by the interaction effects. Furthermore, we show that upon the strain application the character of the anisotropy changes, which should be also reflected in the character of the Cooper-pairing wave function.  \\

 {\it Model and methods:} 
 The crystal field of the O$^{2-}$ oxygen ions breaks the degeneracy of the 4d states of Ru$^{4+}$  into two subshells, the threefold t$_{2g}$ orbitals and the twofold e$_{g}$ orbitals.
The orbital character of the FS  is dominated by  t$_{2g}$ subshell which has a lower energy because the orbitals lobes point between the oxygen ions in contrast to the e$_{g}$ orbitals.
The system is not particle-hole symmetric and has a relatively low effective bandwidth \cite{Medici,Aichhorn}. In other words, there is one additional electron in the half-filled t$_{2g}$ shell, or four electrons per site\cite{Medici2}.
Recently, there have been detailed first-principles calculations on the electronic structure by self-consistent \emph{GW} calculations\cite{Ryee,Aichhorn}. Moreover, a series of studies investigate the correlation effects of these materials\cite{Tran,Pchelkina,Singh,Pchelkina2}. Later, detailed ARPES measurements\cite{Damascelli}  and de Haas-van Alphen experiments\cite{Mackenzie3,Bergemann} were shown to be consistent with LDA bands\cite{DavidSingh,Oguchi}, renormalized due to correlations.  
\\

As a starting point, we use the effective three orbital model including the t$_{2g}$ 
orbital manifold:
\begin{equation}
H_0(\mathbf{k})=\sum_{j,{\bf k}\sigma} \epsilon_{j}({\bf k}) d^{\dagger}_{j,{\bf k}\sigma} d^{}_{j,{\bf k}\sigma},
\end{equation}		
where  orbital indexes are given by $j=xz,yz,xy$, with spinor 
$
\psi^\dagger(\mathbf{k}\sigma)=(d_{xz,\mathbf{k}\sigma}^\dagger, d_{yz,\mathbf{k}\sigma}^\dagger, d_{xy,\mathbf{k}{\bar \sigma}}^\dagger),
$
where   $\mathbf{k}$ and  $\sigma$  (${\bar \sigma}=- \sigma$) represent momentum and spin, respectively.
The electronic dispersion is defined with the help of tight-binding parametrization 
\begin{align}
\begin{aligned}
&
\epsilon_{xz/yz}({\bf k}) = - 2 t_{1/2}   \cos k_x   -
2 t_{2/1}    \cos k_y;
\\ 
&
\epsilon_{xy}({\bf k})  = - 2
t_3 (\cos k_x + \cos k_y) 
\\ 
& - 4 t_4 \cos k_x \cos k_y  
-2
t_5(\cos 2 k_x + \cos 2 k_y ).
\end{aligned}
\end{align}		
and the hopping parameters ($t_1=88$, $t_2=9$, $t_3=80$, $t_4=40$, $t_5=5$, $\mu=109$ (all in meV)) are fitted to the available ARPES experiments \cite{Zabolotnyy}. 
\begin{figure}[t]
	\centering
		\includegraphics[width=0.48\textwidth] {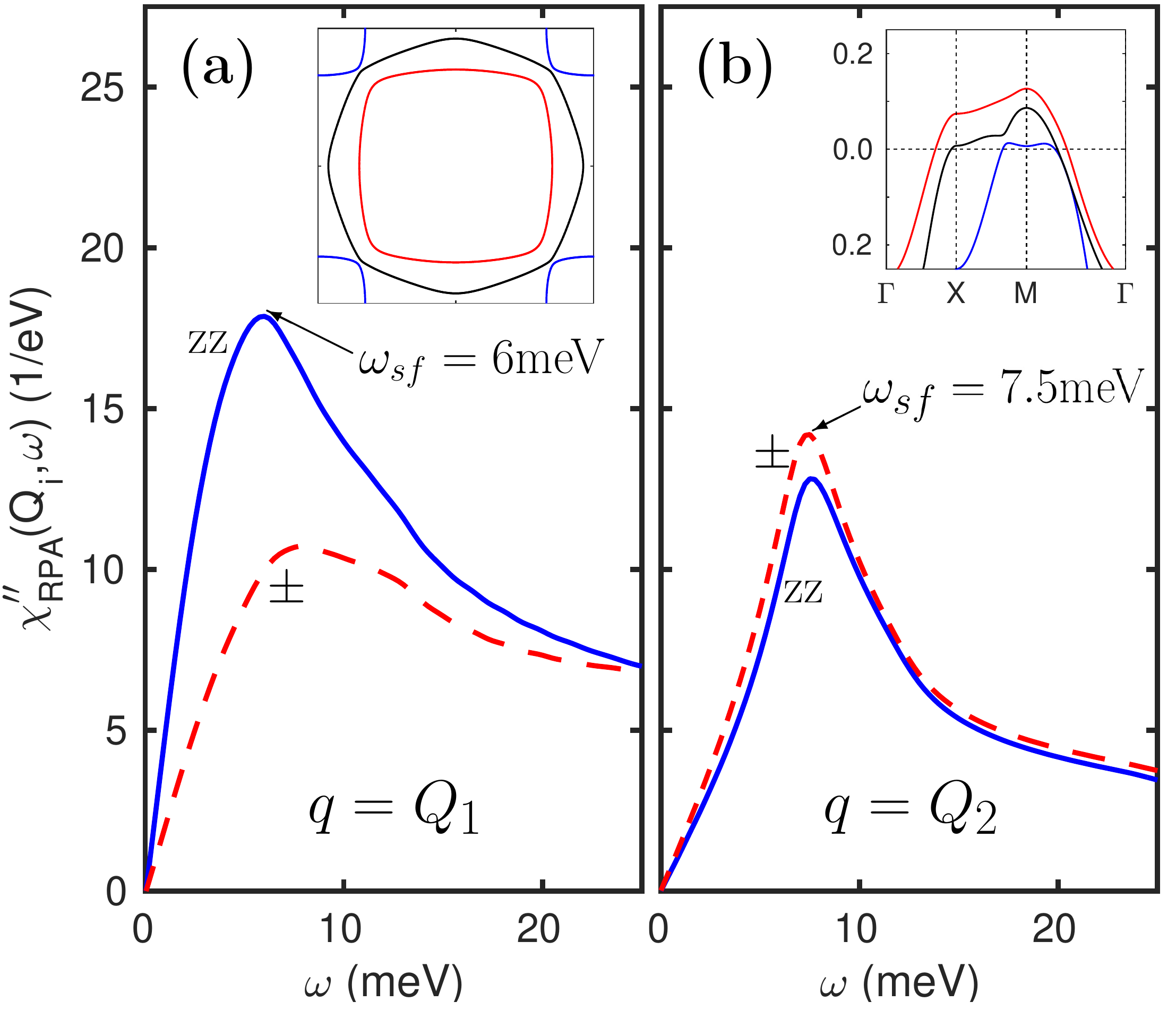}  
	\caption{
	(Colour online)
	Calculated imaginary part of the longitudinal and transverse components of the RPA spin susceptibility at the antiferromagnetic wavevector $\mathbf{Q}_1=(2\pi/3,2\pi/3)$ (a) and $\mathbf{Q}_2=(\pi/6,\pi/6)$ (b). Insets shows the Fermi surface topology the electronic dispersion along high-symmetry lines, respectively.
	}
	\label{Fig:1}
	\label{Fig:2}
\end{figure}
In addition, we  include the on-site spin-orbit coupling\cite{Ng,Eremin2,Simon}, 
$\mathcal{H}_\text{SOC}=\lambda \mathbf{S}\cdot\mathbf{L}$, 
where $\mathbf{S}$ and  $\mathbf{L}$ are the spin and angular momentum operators. Written in terms of the $t_{2g}$ manifold $\bigl(d_{xz\uparrow}^\dagger,d_{yz\uparrow}^\dagger,d_{xy\downarrow}^\dagger, 
d_{xz\downarrow}^\dagger,d_{yz\downarrow}^\dagger,d_{xy\uparrow}^\dagger\bigr)$	the spin-orbit coupling acquires the following form\cite{Eremin2}	
\begin{align}
H_{SOC}&\!=\!\frac{1}{2}\left(\begin{matrix}			
0	& i\lambda  & i\lambda	& 0	& 0	& 0			\\
-i\lambda	& 0 & -\lambda  &	0	& 0	& 0		\\
-i\lambda &	-\lambda	& 0 & 0	& 0	& 0	\\
0	& 0	&	 0   & 0	& -i\lambda	& \lambda			\\
0	& 0	& 0	& i\lambda	& 0	& -i\lambda	\\
0	& 0	& 0	& \lambda	& i\lambda	& 0	\\
\end{matrix}\right),
\end{align}
and we employ $\lambda=35$meV\cite{Zabolotnyy}.
The diagonalization of the combined Hamiltonian, $\mathcal{H}\!=\!\mathcal{H}_0+\mathcal{H}_\text{SOC},$
 yields the electronic band-structure that shows two electron-like Fermi surface (FS) pockets around the $\Gamma$ and a hole-like 
 FS  pocket around the $M$-point of the BZ\cite{Zabolotnyy}. The resulting Fermi surface topology  and band-structure are shown in the insets of Fig.~\ref{Fig:1}(a) and Fig.~\ref{Fig:1}(b), respectively. The interaction part of the Hamiltonian contains the on-site Hubbard-Hund type interactions, written in terms of Hubbard intra- ($U$) and inter-($U'$) orbital terms as well as the residual Hund coupling, $J$.

The physical components of the spin susceptibility are given by
%
\begin{align}
\!\!
\chi^{uv}_0({\mathbf q},{\mi} \Omega)=
\!\!
\frac{-T}{4 N}
\!\!
\sum_{\substack{\mathbf{k}, \mi\omega_n \\ p=q,s=t}}
{\sigma}^{u}_{\gamma\delta} {\sigma}^{v}_{\alpha\beta}
G_{qs,\beta\gamma}^{\mathbf{k},\mi\omega_n} G_{t p, \delta \alpha}^{\mathbf{k'},\mi\omega_n+\mi\Omega},
\end{align}
%
where ${\mathbf k'}=\mathbf{k}+\mathbf{q}$,  and $\sigma^{u=x,y,z}$ are the Pauli matrices. Here, $q,p,s,t$  and $\alpha,\beta,\gamma,\delta$ are the orbital and the spin indexes, respectively. The Green's function is defined by 
$$
G_{ss',\sigma \sigma^\prime}^{\mathbf{k},\mi\omega_n}=
-\int^{\beta}_{0} d \tau e^{\mi \omega \tau} 
\Big\langle T_{\tau} 	
d_{s,\mathbf{k}\sigma}(\tau) d^{\dagger}_{s',\mathbf{k}\sigma^\prime} (0) 
\Big\rangle,
$$
where the transformation from the orbital and the spin basis to the band pseudospin basis is performed by substitution~of
\begin{align}
G_{ss',\sigma\sigma'}^{\mathbf{k},\tau}=\sum_{i} a_{s\sigma}(i,\mathbf{k})a^*_{s' \sigma'}(i,\mathbf{k}) G_{i}(\mathbf{k}, \tau).
\end{align}
Here, $a_{s\sigma}(i,\mathbf{k})$ is the 
 matrix-element that connects band ($i$) 
 and orbital ($s$). 
 Performing the Matsubara frequency sum over~ $\mi \omega_n \longrightarrow \omega+\mi  0^+$,
the expression for the  components of the bare susceptibility 
in the multi-orbital case, is given by
\begin{align}
\!\!
\chi^{uv}_0(\mathbf{q},\omega)=
\sum_{ij,\mathbf{k}}
[\eta^{uv}_{ij;\mathbf{k}\mathbf{k^\prime}}]
\frac{
f(E_{i}^{\mathbf{k}})-f(E_{j}^{\mathbf k'})
}{
E_{j}^{\mathbf k'}-E_{i}^{\mathbf{k}}+\omega+\mi 0^+},
\label{eq:chi}
\end{align}
here the anisotropy of the susceptibility enters through the orbital- and spin-dressing factor
$$
[\eta^{uv}_{ij;\mathbf{k}\mathbf{k^\prime}}]
\!
=
\!
\sigma^{u}_{\alpha\beta}\sigma^{v}_{\gamma\delta}
a_{t\beta}(i,\!\mathbf{k})a^*_{s\gamma}(i,\!\mathbf{k}) %
a_{s\delta}(j,\!\mathbf{k'})a^*_{t\alpha}(j,\!\mathbf{k'})
,
$$
that implies summation over the repeated indexes.

On the diagrammatic level, the bare susceptibility can be also written as:
\begin{align}
\begin{split}
\chi^{xx}_0(\mathbf{q,\omega})&\!=\! %
\diagram{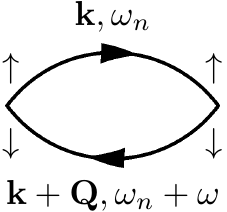}+\diagram{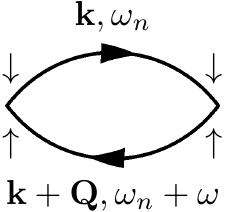}+\diagram{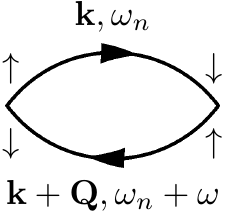}+\diagram{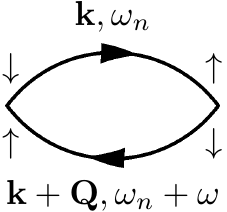}\\
\chi^{yy}_0(\mathbf{q,\omega})&\!=\! %
\diagram{bubble_uudd}+\diagram{bubble_dduu}-\diagram{bubble_udud}-\diagram{bubble_dudu}\\
\chi^{zz}_0(\mathbf{q,\omega})&\!=\! %
\diagram{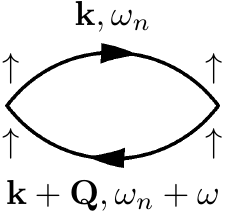}+\diagram{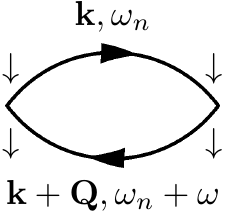}-\diagram{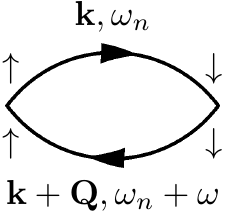}-\diagram{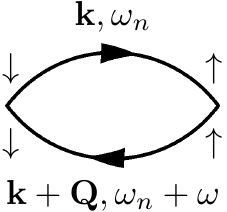}\label{eq:bare_bubbles}
\end{split}
\end{align}
where we define the following notation for the diagrams
\begin{align}
(\csigma)_0&\!=\!\diagramdef{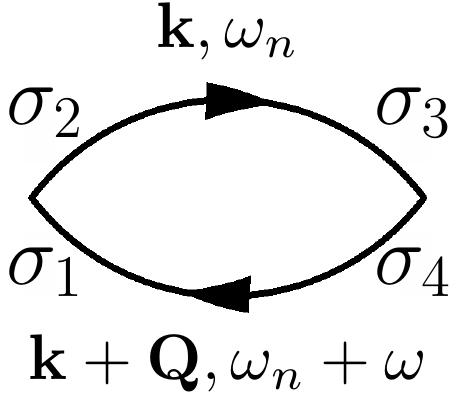}.
\end{align}
This allows to write each components within short-hand notation as 
\begin{align}
\begin{split}
\chi^{xx}_0(\mathbf{q,\omega})&=
\Big(
\cudud+\cdudu+\cuddu+\cduud
\Big)_0,\\
\chi^{yy}_0(\mathbf{q,\omega})&=
\Big(
\cudud+\cdudu-\cuddu-\cduud
\Big)_0,\\
\chi^{zz}_0(\mathbf{q,\omega})&=
\Big(
\cuuuu+\cdddd-\cuudd-\cdduu
\Big)_0.
\end{split}\label{eq:rpa}
\end{align}

Again, for the physical part of the susceptibility, the summation of indexes is implied. Here for zero spin-orbit coupling, $\lambda=0$, the first two bubbles of each component have the same value while the last two bubbles of each component vanish, ensuring the $O(3)$ symmetry of the system.
If spin-orbit coupling acts only among the $d_{xz}$ and $d_{yz}$ orbitals, the only term is $\lambda S_zL_z$ and already leads to a splitting of the transverse and longitudinal part of the susceptibility,
$
\chi^{xx}_0(\mathbf{q,\omega})=\chi^{yy}_0(\mathbf{q,\omega})\neq\chi^{zz}_0(\mathbf{q,\omega}).
$
However, if spin-orbit coupling acts among at least one additional orbital, %
the transverse components $\chi^{xx}_0$ ($\chi^{yy}_0$) differ due to the term $\lambda(S_+L_-+S_-L_+)$, implying full spin anisotropy in the entire BZ
$
\chi^{xx}_0(\mathbf{q,\omega})\neq\chi^{yy}_0(\mathbf{q,\omega})\neq\chi^{zz}_0(\mathbf{q,\omega}).
$
\\

\begin{figure}[b]
	\begin{minipage}{0.485\textwidth}
		\includegraphics[width=\textwidth]{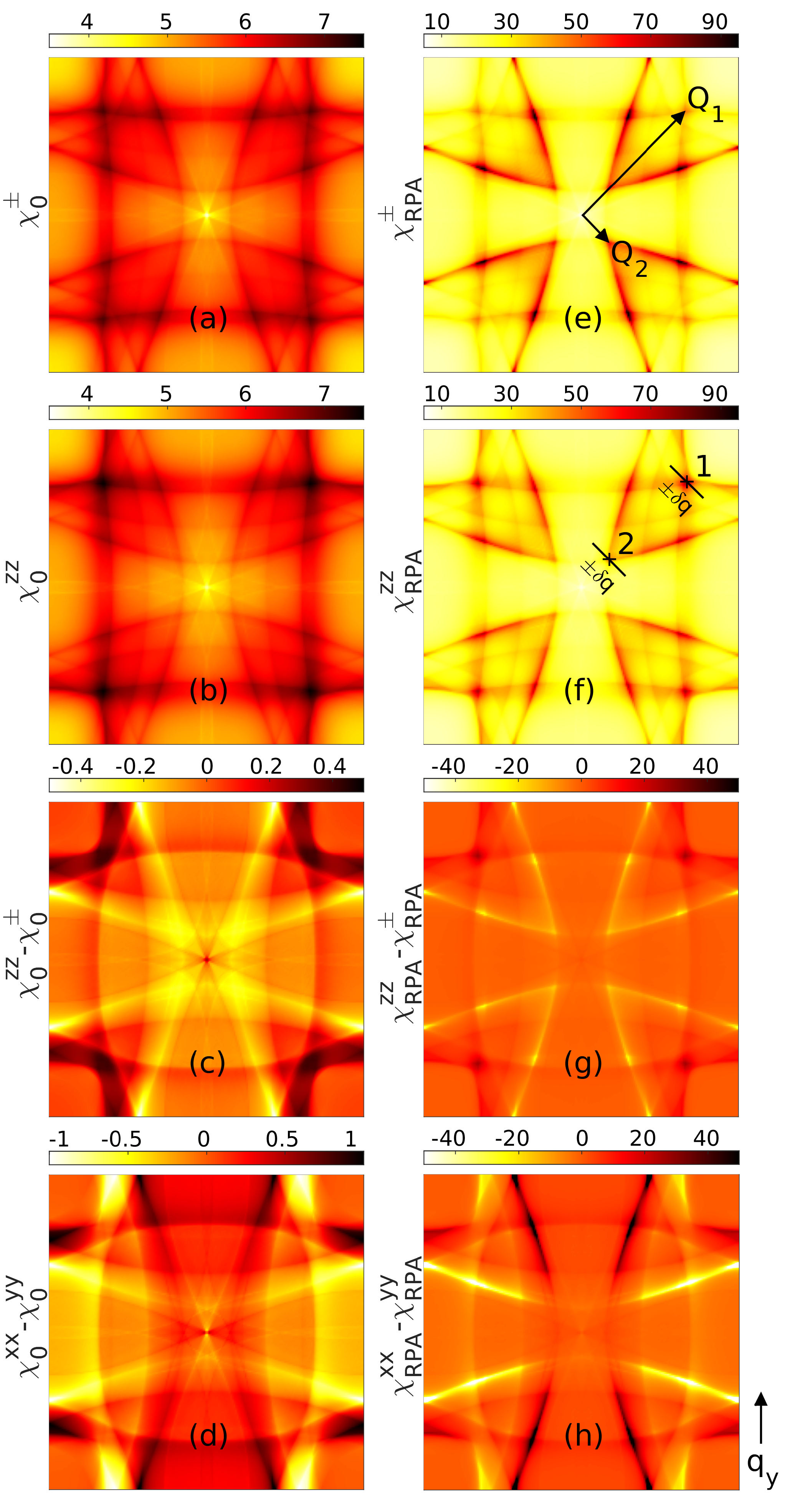}
	\end{minipage}
\caption{ (Colour online)
	Calculated real parts of the bare (a-d) and the RPA (e-h) physical spin susceptibilities as a function of $\mathbf{q}$ in the static limit $\omega=0$ for the longitudinal and transverse components. The units are given in $\pi /a$ in the range $[-1,1]$). }
	\label{Fig:3}
\end{figure}

The diagrammatic treatment for the random phase approximation needs to be done separately for the longitudinal ($zz$) and the transverse ($xx$, $yy$) components of the spin susceptibility. In particular, one finds 
\begin{gather}
\begin{pmatrix}
\nonumber
\cudud	&	\cduud	\\
\cuddu	&	\cdudu	\\
\end{pmatrix}_\text{RPA}=\hspace{0.285\textwidth}\\%
\left[\mathbb{I}-\begin{pmatrix}
\cudud	&	\cduud	\\
\cuddu	&	\cdudu	\\
\end{pmatrix}_0%
\begin{pmatrix}
\Uudud	&	0	\\
0	&	\Ududu	\\
\end{pmatrix}\right]^{-1}%
\begin{pmatrix}
\cudud	&	\cduud	\\
\cuddu	&	\cdudu	\\
\end{pmatrix}_0
,
\nonumber\\
\end{gather}
and
\begin{gather}
\nonumber
\begin{pmatrix}
\cuuuu	&	\cdduu	\\
\cuudd	&	\cdddd	\\
\end{pmatrix}_\text{RPA}=\hspace{0.285\textwidth}\\%
\left[\mathbb{I}-\begin{pmatrix}
\cuuuu	&	\cdduu	\\
\cuudd	&	\cdddd	\\
\end{pmatrix}_0%
\begin{pmatrix}
\Uuuuu	&	\Uuudd	\\
\Udduu	&	\Udddd	\\
\end{pmatrix}\right]^{-1}%
\begin{pmatrix}
\cuuuu	&	\cdduu	\\
\cuudd	&	\cdddd	\\
\end{pmatrix}_0
\nonumber.\\
\end{gather}
Here, each entry of the matrix is a tensor with four orbital indexes $\lbrace pqst\rbrace$, and the summation over orbital indexes for the physical part of the susceptibility has to be performed at the end. Furthermore, 
the matrix equations in the spin space can be decoupled by applying the similarity transformation, 
$S=(
\sigma^{x}
+
\sigma^{z}
)/\sqrt{2}
$,
which yields four decoupled equations. The three equations that correspond to the spin susceptibility are written below,
\begin{equation}
\chi^{uu}_{\text{RPA}}(\mathbf{q},\omega)=
\Big[1-\chi^{uu}_0(\mathbf{q},\omega)U_s
\Big]^{-1}\chi^{uu}_0(\mathbf{q},\omega),
\end{equation}
where 
$uu=xx,yy,zz$.
Furthermore, $U_s\!\equiv\!\Uuuuu\!-\!\Uuudd\!=\!\Uudud$ contains the Hubbard-type on-site interactions $U$, $J$ and $U^\prime\!=\!U\!-\!2J$. In particular, $U$ ($U'$) is the intra- (inter-) orbital Coulomb repulsion, and $J$ represents Hund's coupling. The tensor $U_s$ is given by
\begin{equation}
\begin{aligned}
(U_s)^{aa}_{aa}&=U,& (U_s)^{ab}_{ab}&=U^\prime,
\\
(U_s)^{aa}_{bb}&=J,& (U_s)^{ab}_{ba}&=J.
\end{aligned}
\end{equation}
\\

\begin{figure}[t]
	\begin{minipage}{0.47\textwidth}
		\includegraphics[width=\textwidth]{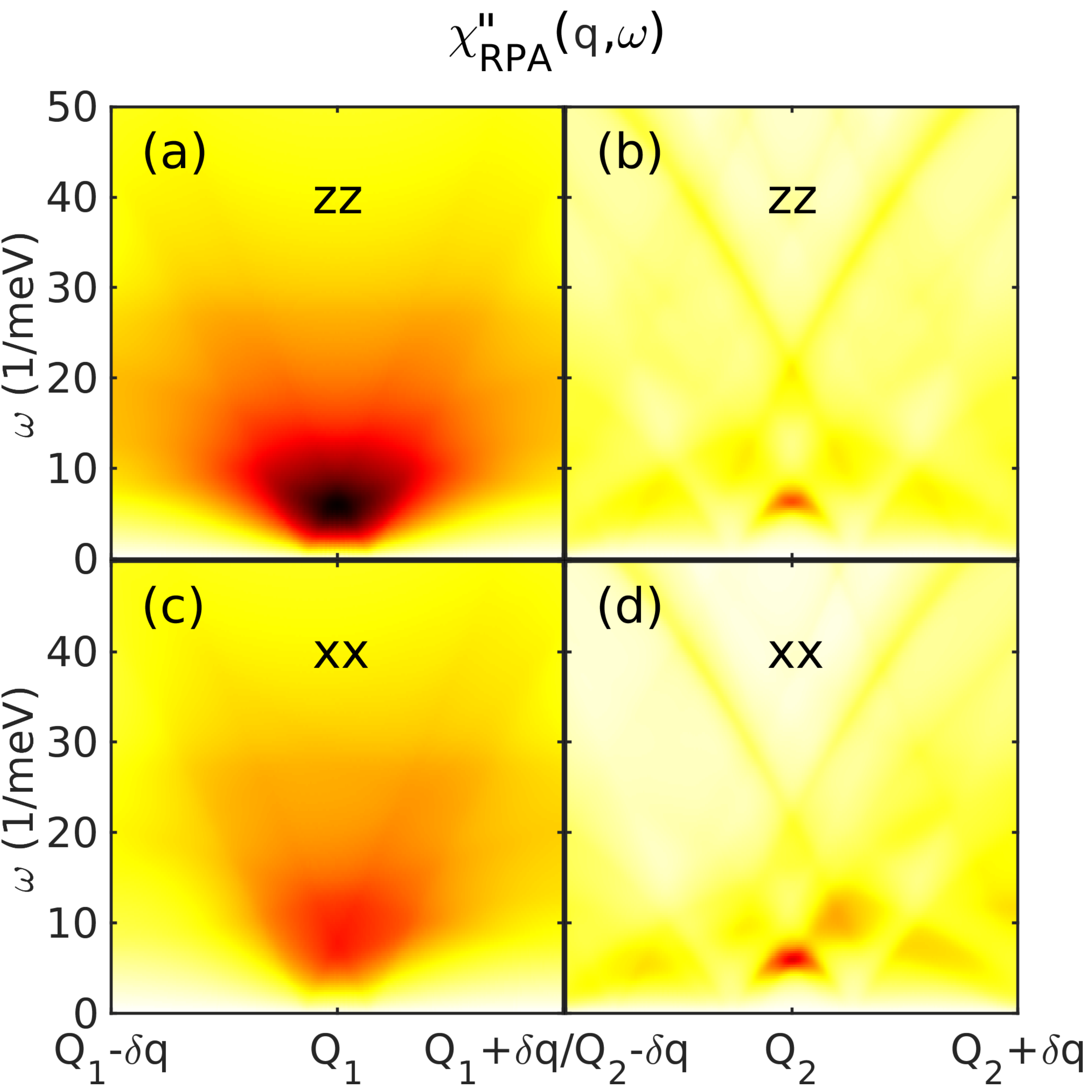}
	\end{minipage}
	\caption{
	(Colour online)
	Calculated frequency and momentum dependencies of the imaginary parts of $\chi^{zz}_\text{RPA}$ and $\chi^{xx}_\text{RPA}$ near $\mathbf{Q}_1$ (a,c) and $\mathbf{Q}_2$ (b,d).  
	The two straight lines defined by the endpoints of $(\mathbf{Q}_i-\delta q,\mathbf{Q}_i+\delta q)$, $i\!=\!1,2$, are visualised in Fig. 2(f).
	}
	\label{Fig:4}
\end{figure}
%

{\it Numerical results:} 
The well-known fact of the electronic structure of the Sr$_2$RuO$_4$ is the nesting of the quasi-one dimensional $xz$, and $yz$-bands at the incommensurate wave vector ${\bf Q}_1$\cite{Eremin}. In the inelastic neutron scattering this nesting yields the incommensurate magnetic fluctuations peaked at $\omega_{sf}=6$meV, which are polarized along $z$-direction. The parameters of the non-interacting Hamiltonian are fixed in our case by the fit to the ARPES experiments\cite{Zabolotnyy}. Thus we employ $U=0.12$eV and $J=0.25U$ to reproduce the frequency position of 6meV of the incommensurate spin fluctuations at ${\bf Q}_1$ in the longitudinal response, as shown in Fig.1(a).  At the next step we find that the transverse fluctuations are peaked at more or less the same frequency but appear to be with factor 2 smaller intensity, which again agrees very well with the neutron scattering data\cite{Sidis}. Analysing each component of the RPA susceptibility in detail we find that the easy-axis ($z$) polarization of the incommensurate AF fluctuation at {\bf Q}$_1$ occurs due to the dominant interband nesting of the $xz$ ($yz$) bands. 

As it is generally believed that the pure AF fluctuation cannot be responsible for the triplet character of the Cooper-pairing, we have analysed the behaviour of the spin response in the entire BZ. In particular, in Fig. 2 we show the results of the RPA physical susceptibility and its anisotropy in the first BZ. In addition to the incommensurate AF fluctuations at {\bf Q}$_1$ we also find dispersing magnetic excitation, peaked at much smaller momentum  $\mathbf{Q}_2=(\pi/6,\pi/6)$. These small {\bf q} excitations originate mostly from the $xy$-band and were also observed previously experimentally as a quite broad feature\cite{Braden2}. As a consequence of this, they are weakly anisotropic,  as shown in Fig.1(b) and for the value of the interactions employed are peaked at energies of about 7.5meV. Surprisingly their intensity appears to have similar magnitude as the excitations at {\bf Q}$_1$ and therefore they may play an important role for the Cooper-pairing. 

Another interesting feature we see from Fig.~\ref{Fig:3} is that maximal magnetic anisotropy of the spin fluctuations in Sr$_2$RuO$_4$ away from the high symmetry directions. The components of the susceptibility remain anisotropic and in general  away from any high-symmetry points one observes $\chi^{xx}\neq\chi^{yy}\neq\chi^{zz}$. 
In particular, $\chi^{zz}>\chi^{xx/yy}$ near the M-point, 
yet  $\chi^{xx}<\chi^{yy}$ and $\chi^{xx}>\chi^{yy}$ around X-, and Y-points, respectively, implying a breaking of the in-plane symmetry of the spin susceptibility. Note that such an anisotropy of the spin fluctuations is related to the spin-orbit coupling that transfers the highly anisotropic orbital character of the $xz$, and $yz$-orbitals to the spin subspace. Theses anisotropies should be maximally seen in the dispersion of the both excitations at ${\bf Q}_1$ and {\bf Q}$_2$, which we present in Fig.\ref{Fig:4}.
\\

Observe that this anisotropy should become strongly visible once the strain is applied. In the following we include its effect on the electronic structure via anisotropic intra-orbital hopping parameters along $x$ and $y$ direction of the kinetic part of the Hamiltonian,  such that it breaks the $C_4$ symmetry of the system, similar to Ref.\onlinecite{Steppke}. We find that for $2.65\%$ strain the Fermi surface of the $xy$-band touches the Van~Hove~singularity at the Y point.

Although all FS  pockets are just $C_2$ symmetric under strain, the larger electron band around $\Gamma$ is distorted significantly\cite{Liu} and is responsible for the sharp increase of states at the Fermi level which can be seen in Fig.~\ref{Fig:4}(a). 
For strain values close to the Van~Hove~singularity, we find that the dominating peak of the real part of the magnetic response shifts from {\bf Q}$_2$ to the ferromagnetic ones ${\bf q} =0$ and has an Ising like character.  
As a consequence, the Ising ferromagnetic instability at $\mathbf{q}=(0,0)$ is much larger than the one for $\mathbf{Q}_1$ and $\mathbf{Q}_2$, implying that the magnetic moments point towards the $y$-direction. 
\\

\begin{figure}
	\begin{minipage}{0.485\textwidth}
		\includegraphics[width=\textwidth,trim=0mm 0mm 0mm 0mm,clip]{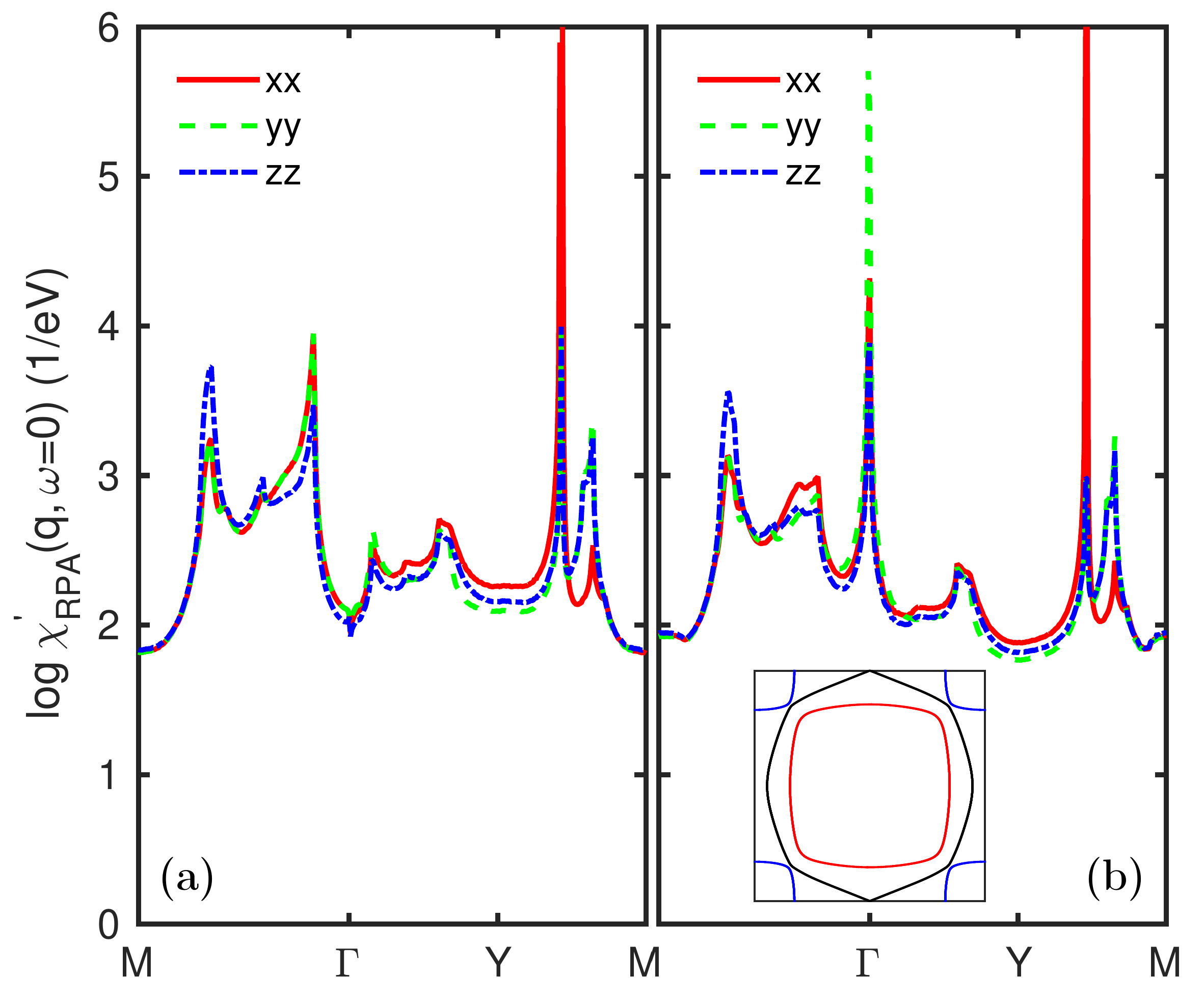}
	\end{minipage}
	\caption{
	(Colour online) Calculated real part of the RPA spin susceptibility along high-symmetry directions of the BZ. 
	The anisotropy of the longitudinal and transverse components results from spin-orbit coupling (a);
	and results from spin-orbit coupling including anisotropic strain (b); the anisotropy at $\Gamma$ %
	is $yy > xx > zz$.
	 Inset shows the Fermi surface topology with Van~Hove~singularity at Y-point.}
	 \label{Fig:4} 
\end{figure}

{\it Conclusion:} To conclude we study the anisotropy of the spin fluctuations in Sr$_2$RuO$4$ in the presence of spin-orbit coupling and anisotropic strain using quasi-two-dimensional tight-binding parametrization fitted to the ARPES results.
Similar to the previous observations we find the in-plane polarization of the low ${\bf q}$ magnetic fluctuations and the out-of-plane polarization of the incommensurate magnetic fluctuation at the nesting wave vector {\bf Q}$_1 = (2/3 \pi ,2/3 \pi)$. Most importantly we also find strong fluctuations near much smaller wave vector {\bf Q}$_2=(\pi/6,\pi/6)$, which shows very weak anisotropy. Furthermore, one finds that apart from the high-symmetry direction of the tetragonal Brillouin zone the magnetic anisotropy is maximal, i.e. $\chi^{xx} \neq \chi^{yy} \neq \chi^{zz}$. This is the consequence of the orbital anisotropy of the $xz$ and $yz$ orbitals in the momentum space. We also study how the magnetic anisotropy evolves in the presence of the strain and finds strong Ising-like ferromagnetic fluctuations which appear when the $xy$-band touches the Van Hove point at the $Y$-point of the BZ.
\\

{\it Acknowledgments:}
We acknowledge helpful discussions with B. Andersen, M. Braden, M.M. Korshunov, P. Hirschfeld, A. Romer, and P. Thalmeier.
S.C. and  A.A. wish to acknowledge the Korea Ministry of Education, Science and Technology, Gyeongsangbuk-Do and Pohang City for Independent Junior Research Groups at the Asia Pacific Center for Theoretical Physics. 
The work by  A.A. was supported through  NRF funded by MSIP of Korea (2015R1C1A1A01052411), and by  Max Planck POSTECH / KOREA Research Initiative (No. 2011-0031558) programs through NRF funded by MSIP of Korea.

\bibliographystyle{apsrev4-1}
\bibliography{sr} 

\begin{thebibliography}{52}%
\makeatletter
\providecommand \@ifxundefined [1]{%
 \@ifx{#1\undefined}
}%
\providecommand \@ifnum [1]{%
 \ifnum #1\expandafter \@firstoftwo
 \else \expandafter \@secondoftwo
 \fi
}%
\providecommand \@ifx [1]{%
 \ifx #1\expandafter \@firstoftwo
 \else \expandafter \@secondoftwo
 \fi
}%
\providecommand \natexlab [1]{#1}%
\providecommand \enquote  [1]{``#1''}%
\providecommand \bibnamefont  [1]{#1}%
\providecommand \bibfnamefont [1]{#1}%
\providecommand \citenamefont [1]{#1}%
\providecommand \href@noop [0]{\@secondoftwo}%
\providecommand \href [0]{\begingroup \@sanitize@url \@href}%
\providecommand \@href[1]{\@@startlink{#1}\@@href}%
\providecommand \@@href[1]{\endgroup#1\@@endlink}%
\providecommand \@sanitize@url [0]{\catcode `\\12\catcode `\$12\catcode
  `\&12\catcode `\#12\catcode `\^12\catcode `\_12\catcode `\%12\relax}%
\providecommand \@@startlink[1]{}%
\providecommand \@@endlink[0]{}%
\providecommand \url  [0]{\begingroup\@sanitize@url \@url }%
\providecommand \@url [1]{\endgroup\@href {#1}{\urlprefix }}%
\providecommand \urlprefix  [0]{URL }%
\providecommand \Eprint [0]{\href }%
\providecommand \doibase [0]{http://dx.doi.org/}%
\providecommand \selectlanguage [0]{\@gobble}%
\providecommand \bibinfo  [0]{\@secondoftwo}%
\providecommand \bibfield  [0]{\@secondoftwo}%
\providecommand \translation [1]{[#1]}%
\providecommand \BibitemOpen [0]{}%
\providecommand \bibitemStop [0]{}%
\providecommand \bibitemNoStop [0]{.\EOS\space}%
\providecommand \EOS [0]{\spacefactor3000\relax}%
\providecommand \BibitemShut  [1]{\csname bibitem#1\endcsname}%
\let\auto@bib@innerbib\@empty
\bibitem [{\citenamefont {Maeno}\ \emph {et~al.}(1994)\citenamefont {Maeno},
  \citenamefont {Hashimoto}, \citenamefont {Yoshida}, \citenamefont
  {Nishizaki}, \citenamefont {Fujita}, \citenamefont {Bednorz},\ and\
  \citenamefont {Lichtenberg}}]{Maeno}%
  \BibitemOpen
  \bibfield  {author} {\bibinfo {author} {\bibfnamefont {Y.}~\bibnamefont
  {Maeno}}, \bibinfo {author} {\bibfnamefont {H.}~\bibnamefont {Hashimoto}},
  \bibinfo {author} {\bibfnamefont {K.}~\bibnamefont {Yoshida}}, \bibinfo
  {author} {\bibfnamefont {S.}~\bibnamefont {Nishizaki}}, \bibinfo {author}
  {\bibfnamefont {T.}~\bibnamefont {Fujita}}, \bibinfo {author} {\bibfnamefont
  {J.~G.}\ \bibnamefont {Bednorz}}, \ and\ \bibinfo {author} {\bibfnamefont
  {F.}~\bibnamefont {Lichtenberg}},\ }\href
  {http://dx.doi.org/10.1038/372532a0} {\bibfield  {journal} {\bibinfo
  {journal} {Nature}\ }\textbf {\bibinfo {volume} {372}},\ \bibinfo {pages}
  {532} (\bibinfo {year} {1994})}\BibitemShut {NoStop}%
\bibitem [{\citenamefont {Mackenzie}\ and\ \citenamefont
  {Maeno}(2003)}]{Mackenzie}%
  \BibitemOpen
  \bibfield  {author} {\bibinfo {author} {\bibfnamefont {A.~P.}\ \bibnamefont
  {Mackenzie}}\ and\ \bibinfo {author} {\bibfnamefont {Y.}~\bibnamefont
  {Maeno}},\ }\href {http://link.aps.org/doi/10.1103/RevModPhys.75.657}
  {\bibfield  {journal} {\bibinfo  {journal} {Rev. Mod. Phys.}\ }\textbf
  {\bibinfo {volume} {75}},\ \bibinfo {pages} {657} (\bibinfo {year}
  {2003})}\BibitemShut {NoStop}%
\bibitem [{\citenamefont {Ishida}\ \emph {et~al.}(1998)\citenamefont {Ishida},
  \citenamefont {Mukuda}, \citenamefont {Kitaoka}, \citenamefont {Asayama},
  \citenamefont {Mao}, \citenamefont {Mori},\ and\ \citenamefont
  {Maeno}}]{Ishida}%
  \BibitemOpen
  \bibfield  {author} {\bibinfo {author} {\bibfnamefont {K.}~\bibnamefont
  {Ishida}}, \bibinfo {author} {\bibfnamefont {H.}~\bibnamefont {Mukuda}},
  \bibinfo {author} {\bibfnamefont {Y.}~\bibnamefont {Kitaoka}}, \bibinfo
  {author} {\bibfnamefont {K.}~\bibnamefont {Asayama}}, \bibinfo {author}
  {\bibfnamefont {Z.~Q.}\ \bibnamefont {Mao}}, \bibinfo {author} {\bibfnamefont
  {Y.}~\bibnamefont {Mori}}, \ and\ \bibinfo {author} {\bibfnamefont
  {Y.}~\bibnamefont {Maeno}},\ }\href {http://dx.doi.org/10.1038/25315}
  {\bibfield  {journal} {\bibinfo  {journal} {Nature}\ }\textbf {\bibinfo
  {volume} {396}},\ \bibinfo {pages} {658} (\bibinfo {year}
  {1998})}\BibitemShut {NoStop}%
\bibitem [{\citenamefont {Ishida}\ \emph {et~al.}(2015)\citenamefont {Ishida},
  \citenamefont {Manago}, \citenamefont {Yamanaka}, \citenamefont {Fukazawa},
  \citenamefont {Mao}, \citenamefont {Maeno},\ and\ \citenamefont
  {Miyake}}]{Ishida2}%
  \BibitemOpen
  \bibfield  {author} {\bibinfo {author} {\bibfnamefont {K.}~\bibnamefont
  {Ishida}}, \bibinfo {author} {\bibfnamefont {M.}~\bibnamefont {Manago}},
  \bibinfo {author} {\bibfnamefont {T.}~\bibnamefont {Yamanaka}}, \bibinfo
  {author} {\bibfnamefont {H.}~\bibnamefont {Fukazawa}}, \bibinfo {author}
  {\bibfnamefont {Z.~Q.}\ \bibnamefont {Mao}}, \bibinfo {author} {\bibfnamefont
  {Y.}~\bibnamefont {Maeno}}, \ and\ \bibinfo {author} {\bibfnamefont
  {K.}~\bibnamefont {Miyake}},\ }\href {\doibase 10.1103/PhysRevB.92.100502}
  {\bibfield  {journal} {\bibinfo  {journal} {Phys. Rev. B}\ }\textbf {\bibinfo
  {volume} {92}},\ \bibinfo {pages} {100502} (\bibinfo {year}
  {2015})}\BibitemShut {NoStop}%
\bibitem [{\citenamefont {Duffy}\ \emph {et~al.}(2000)\citenamefont {Duffy},
  \citenamefont {Hayden}, \citenamefont {Maeno}, \citenamefont {Mao},
  \citenamefont {Kulda},\ and\ \citenamefont {McIntyre}}]{Duffy}%
  \BibitemOpen
  \bibfield  {author} {\bibinfo {author} {\bibfnamefont {J.~A.}\ \bibnamefont
  {Duffy}}, \bibinfo {author} {\bibfnamefont {S.~M.}\ \bibnamefont {Hayden}},
  \bibinfo {author} {\bibfnamefont {Y.}~\bibnamefont {Maeno}}, \bibinfo
  {author} {\bibfnamefont {Z.}~\bibnamefont {Mao}}, \bibinfo {author}
  {\bibfnamefont {J.}~\bibnamefont {Kulda}}, \ and\ \bibinfo {author}
  {\bibfnamefont {G.~J.}\ \bibnamefont {McIntyre}},\ }\href {\doibase
  10.1103/PhysRevLett.85.5412} {\bibfield  {journal} {\bibinfo  {journal}
  {Phys. Rev. Lett.}\ }\textbf {\bibinfo {volume} {85}},\ \bibinfo {pages}
  {5412} (\bibinfo {year} {2000})}\BibitemShut {NoStop}%
\bibitem [{\citenamefont {Xia}\ \emph {et~al.}(2006)\citenamefont {Xia},
  \citenamefont {Maeno}, \citenamefont {Beyersdorf}, \citenamefont {Fejer},\
  and\ \citenamefont {Kapitulnik}}]{Xia}%
  \BibitemOpen
  \bibfield  {author} {\bibinfo {author} {\bibfnamefont {J.}~\bibnamefont
  {Xia}}, \bibinfo {author} {\bibfnamefont {Y.}~\bibnamefont {Maeno}}, \bibinfo
  {author} {\bibfnamefont {P.~T.}\ \bibnamefont {Beyersdorf}}, \bibinfo
  {author} {\bibfnamefont {M.~M.}\ \bibnamefont {Fejer}}, \ and\ \bibinfo
  {author} {\bibfnamefont {A.}~\bibnamefont {Kapitulnik}},\ }\href {\doibase
  10.1103/PhysRevLett.97.167002} {\bibfield  {journal} {\bibinfo  {journal}
  {Phys. Rev. Lett.}\ }\textbf {\bibinfo {volume} {97}},\ \bibinfo {pages}
  {167002} (\bibinfo {year} {2006})}\BibitemShut {NoStop}%
\bibitem [{\citenamefont {Maeno}\ \emph {et~al.}(2011)\citenamefont {Maeno},
  \citenamefont {Kittaka}, \citenamefont {Nomura}, \citenamefont {Yonezawa},\
  and\ \citenamefont {Ishida}}]{Maeno2}%
  \BibitemOpen
  \bibfield  {author} {\bibinfo {author} {\bibfnamefont {Y.}~\bibnamefont
  {Maeno}}, \bibinfo {author} {\bibfnamefont {S.}~\bibnamefont {Kittaka}},
  \bibinfo {author} {\bibfnamefont {T.}~\bibnamefont {Nomura}}, \bibinfo
  {author} {\bibfnamefont {S.}~\bibnamefont {Yonezawa}}, \ and\ \bibinfo
  {author} {\bibfnamefont {K.}~\bibnamefont {Ishida}},\ }\href {\doibase
  10.1143/JPSJ.81.011009} {\bibfield  {journal} {\bibinfo  {journal} {J. Phys.
  Soc. Jpn.}\ }\textbf {\bibinfo {volume} {81}},\ \bibinfo {pages} {011009}
  (\bibinfo {year} {2011})}\BibitemShut {NoStop}%
\bibitem [{\citenamefont {Kidwingira}\ \emph {et~al.}(2006)\citenamefont
  {Kidwingira}, \citenamefont {Strand}, \citenamefont {Van~Harlingen},\ and\
  \citenamefont {Maeno}}]{Kidwingira}%
  \BibitemOpen
  \bibfield  {author} {\bibinfo {author} {\bibfnamefont {F.}~\bibnamefont
  {Kidwingira}}, \bibinfo {author} {\bibfnamefont {J.~D.}\ \bibnamefont
  {Strand}}, \bibinfo {author} {\bibfnamefont {D.~J.}\ \bibnamefont
  {Van~Harlingen}}, \ and\ \bibinfo {author} {\bibfnamefont {Y.}~\bibnamefont
  {Maeno}},\ }\href {\doibase 10.1126/science.1133239} {\bibfield  {journal}
  {\bibinfo  {journal} {Science}\ }\textbf {\bibinfo {volume} {314}},\ \bibinfo
  {pages} {1267} (\bibinfo {year} {2006})}\BibitemShut {NoStop}%
\bibitem [{\citenamefont {Nelson}\ \emph {et~al.}(2004)\citenamefont {Nelson},
  \citenamefont {Mao}, \citenamefont {Maeno},\ and\ \citenamefont
  {Liu}}]{Nelson}%
  \BibitemOpen
  \bibfield  {author} {\bibinfo {author} {\bibfnamefont {K.~D.}\ \bibnamefont
  {Nelson}}, \bibinfo {author} {\bibfnamefont {Z.~Q.}\ \bibnamefont {Mao}},
  \bibinfo {author} {\bibfnamefont {Y.}~\bibnamefont {Maeno}}, \ and\ \bibinfo
  {author} {\bibfnamefont {Y.}~\bibnamefont {Liu}},\ }\href {\doibase
  10.1126/science.1103881} {\bibfield  {journal} {\bibinfo  {journal}
  {Science}\ }\textbf {\bibinfo {volume} {306}},\ \bibinfo {pages} {1151}
  (\bibinfo {year} {2004})}\BibitemShut {NoStop}%
\bibitem [{\citenamefont {Wang}\ \emph
  {et~al.}(2013{\natexlab{a}})\citenamefont {Wang}, \citenamefont {Platt},
  \citenamefont {Yang}, \citenamefont {Honerkamp}, \citenamefont {Zhang},
  \citenamefont {Hanke}, \citenamefont {Rice},\ and\ \citenamefont
  {Thomale}}]{Thomale}%
  \BibitemOpen
  \bibfield  {author} {\bibinfo {author} {\bibfnamefont {Q.~H.}\ \bibnamefont
  {Wang}}, \bibinfo {author} {\bibfnamefont {C.}~\bibnamefont {Platt}},
  \bibinfo {author} {\bibfnamefont {Y.}~\bibnamefont {Yang}}, \bibinfo {author}
  {\bibfnamefont {C.}~\bibnamefont {Honerkamp}}, \bibinfo {author}
  {\bibfnamefont {F.~C.}\ \bibnamefont {Zhang}}, \bibinfo {author}
  {\bibfnamefont {W.}~\bibnamefont {Hanke}}, \bibinfo {author} {\bibfnamefont
  {T.~M.}\ \bibnamefont {Rice}}, \ and\ \bibinfo {author} {\bibfnamefont
  {R.}~\bibnamefont {Thomale}},\ }\href
  {http://stacks.iop.org/0295-5075/104/i=1/a=17013} {\bibfield  {journal}
  {\bibinfo  {journal} {Europhysics Letters)}\ }\textbf {\bibinfo {volume}
  {104}},\ \bibinfo {pages} {17013} (\bibinfo {year}
  {2013}{\natexlab{a}})}\BibitemShut {NoStop}%
\bibitem [{\citenamefont {Huo}\ and\ \citenamefont {Zhang}(2013)}]{Wei}%
  \BibitemOpen
  \bibfield  {author} {\bibinfo {author} {\bibfnamefont {J.-W.}\ \bibnamefont
  {Huo}}\ and\ \bibinfo {author} {\bibfnamefont {F.-C.}\ \bibnamefont
  {Zhang}},\ }\href {\doibase 10.1103/PhysRevB.87.134501} {\bibfield  {journal}
  {\bibinfo  {journal} {Phys. Rev. B}\ }\textbf {\bibinfo {volume} {87}},\
  \bibinfo {pages} {134501} (\bibinfo {year} {2013})}\BibitemShut {NoStop}%
\bibitem [{\citenamefont {Rastovski}\ \emph {et~al.}(2013)\citenamefont
  {Rastovski}, \citenamefont {Dewhurst}, \citenamefont {Gannon}, \citenamefont
  {Peets}, \citenamefont {Takatsu}, \citenamefont {Maeno}, \citenamefont
  {Ichioka}, \citenamefont {Machida},\ and\ \citenamefont
  {Eskildsen}}]{Rastovski}%
  \BibitemOpen
  \bibfield  {author} {\bibinfo {author} {\bibfnamefont {C.}~\bibnamefont
  {Rastovski}}, \bibinfo {author} {\bibfnamefont {C.~D.}\ \bibnamefont
  {Dewhurst}}, \bibinfo {author} {\bibfnamefont {W.~J.}\ \bibnamefont
  {Gannon}}, \bibinfo {author} {\bibfnamefont {D.~C.}\ \bibnamefont {Peets}},
  \bibinfo {author} {\bibfnamefont {H.}~\bibnamefont {Takatsu}}, \bibinfo
  {author} {\bibfnamefont {Y.}~\bibnamefont {Maeno}}, \bibinfo {author}
  {\bibfnamefont {M.}~\bibnamefont {Ichioka}}, \bibinfo {author} {\bibfnamefont
  {K.}~\bibnamefont {Machida}}, \ and\ \bibinfo {author} {\bibfnamefont
  {M.~R.}\ \bibnamefont {Eskildsen}},\ }\href {\doibase
  10.1103/PhysRevLett.111.087003} {\bibfield  {journal} {\bibinfo  {journal}
  {Phys. Rev. Lett.}\ }\textbf {\bibinfo {volume} {111}},\ \bibinfo {pages}
  {087003} (\bibinfo {year} {2013})}\BibitemShut {NoStop}%
\bibitem [{\citenamefont {Yoshiteru~Maeno}(2001)}]{Maeno3}%
  \BibitemOpen
  \bibfield  {author} {\bibinfo {author} {\bibfnamefont {M.~S.}\ \bibnamefont
  {Yoshiteru~Maeno}, \bibfnamefont {T.~Maurice~Rice}},\ }\href
  {http://dx.doi.org/10.1063/1.1349611} {\bibfield  {journal} {\bibinfo
  {journal} {Physics Today}\ }\textbf {\bibinfo {volume} {54}},\ \bibinfo
  {pages} {42} (\bibinfo {year} {2001})}\BibitemShut {NoStop}%
\bibitem [{\citenamefont {Akbari}\ and\ \citenamefont
  {Thalmeier}(2013)}]{Akbari}%
  \BibitemOpen
  \bibfield  {author} {\bibinfo {author} {\bibfnamefont {A.}~\bibnamefont
  {Akbari}}\ and\ \bibinfo {author} {\bibfnamefont {P.}~\bibnamefont
  {Thalmeier}},\ }\href {\doibase 10.1103/PhysRevB.88.134519} {\bibfield
  {journal} {\bibinfo  {journal} {Phys. Rev. B}\ }\textbf {\bibinfo {volume}
  {88}},\ \bibinfo {pages} {134519} (\bibinfo {year} {2013})}\BibitemShut
  {NoStop}%
\bibitem [{\citenamefont {Annett}\ \emph {et~al.}(2006)\citenamefont {Annett},
  \citenamefont {Litak}, \citenamefont {Gy\"orffy},\ and\ \citenamefont
  {Wysoki\ifmmode~\acute{n}\else \'{n}\fi{}ski}}]{Annett}%
  \BibitemOpen
  \bibfield  {author} {\bibinfo {author} {\bibfnamefont {J.~F.}\ \bibnamefont
  {Annett}}, \bibinfo {author} {\bibfnamefont {G.}~\bibnamefont {Litak}},
  \bibinfo {author} {\bibfnamefont {B.~L.}\ \bibnamefont {Gy\"orffy}}, \ and\
  \bibinfo {author} {\bibfnamefont {K.~I.}\ \bibnamefont
  {Wysoki\ifmmode~\acute{n}\else \'{n}\fi{}ski}},\ }\href {\doibase
  10.1103/PhysRevB.73.134501} {\bibfield  {journal} {\bibinfo  {journal} {Phys.
  Rev. B}\ }\textbf {\bibinfo {volume} {73}},\ \bibinfo {pages} {134501}
  (\bibinfo {year} {2006})}\BibitemShut {NoStop}%
\bibitem [{\citenamefont {Eremin}\ \emph {et~al.}(2004)\citenamefont {Eremin},
  \citenamefont {Manske}, \citenamefont {Ovchinnikov},\ and\ \citenamefont
  {Annett}}]{Eremin}%
  \BibitemOpen
  \bibfield  {author} {\bibinfo {author} {\bibfnamefont {I.}~\bibnamefont
  {Eremin}}, \bibinfo {author} {\bibfnamefont {D.}~\bibnamefont {Manske}},
  \bibinfo {author} {\bibfnamefont {S.}~\bibnamefont {Ovchinnikov}}, \ and\
  \bibinfo {author} {\bibfnamefont {J.}~\bibnamefont {Annett}},\ }\href
  {http://dx.doi.org/10.1002/andp.200310069} {\bibfield  {journal} {\bibinfo
  {journal} {Ann. Phys.}\ }\textbf {\bibinfo {volume} {13}},\ \bibinfo {pages}
  {149} (\bibinfo {year} {2004})}\BibitemShut {NoStop}%
\bibitem [{\citenamefont {Raghu}\ \emph {et~al.}(2010)\citenamefont {Raghu},
  \citenamefont {Kapitulnik},\ and\ \citenamefont {Kivelson}}]{Raghu:2010aa}%
  \BibitemOpen
  \bibfield  {author} {\bibinfo {author} {\bibfnamefont {S.}~\bibnamefont
  {Raghu}}, \bibinfo {author} {\bibfnamefont {A.}~\bibnamefont {Kapitulnik}}, \
  and\ \bibinfo {author} {\bibfnamefont {S.~A.}\ \bibnamefont {Kivelson}},\
  }\href {\doibase 10.1103/PhysRevLett.105.136401} {\bibfield  {journal}
  {\bibinfo  {journal} {Phys. Rev. Lett.}\ }\textbf {\bibinfo {volume} {105}},\
  \bibinfo {pages} {136401} (\bibinfo {year} {2010})}\BibitemShut {NoStop}%
\bibitem [{\citenamefont {Masahisa~Tsuchiizu}\ and\ \citenamefont
  {Kontani}(2015)}]{Tsuchiizu2015}%
  \BibitemOpen
  \bibfield  {author} {\bibinfo {author} {\bibfnamefont {S.~O. Y.~O.}\
  \bibnamefont {Masahisa~Tsuchiizu}, \bibfnamefont {Youichi~Yamakawa}}\ and\
  \bibinfo {author} {\bibfnamefont {H.}~\bibnamefont {Kontani}},\ }\href
  {\doibase 10.1103/PhysRevB.91.155103} {\bibfield  {journal} {\bibinfo
  {journal} {Phys. Rev. B}\ }\textbf {\bibinfo {volume} {91}},\ \bibinfo
  {pages} {155103} (\bibinfo {year} {2015})}\BibitemShut {NoStop}%
\bibitem [{\citenamefont {Taniguchi}\ \emph {et~al.}(2014)\citenamefont
  {Taniguchi}, \citenamefont {Nishimura}, \citenamefont {Goh}, \citenamefont
  {Yonezawa},\ and\ \citenamefont {Maeno}}]{Taniguchi}%
  \BibitemOpen
  \bibfield  {author} {\bibinfo {author} {\bibfnamefont {H.}~\bibnamefont
  {Taniguchi}}, \bibinfo {author} {\bibfnamefont {K.}~\bibnamefont
  {Nishimura}}, \bibinfo {author} {\bibfnamefont {S.~K.}\ \bibnamefont {Goh}},
  \bibinfo {author} {\bibfnamefont {S.}~\bibnamefont {Yonezawa}}, \ and\
  \bibinfo {author} {\bibfnamefont {Y.}~\bibnamefont {Maeno}},\ }\bibfield
  {booktitle} {\emph {\bibinfo {booktitle} {Journal of the Physical Society of
  Japan}},\ }\href {\doibase 10.7566/JPSJ.84.014707} {\bibfield  {journal}
  {\bibinfo  {journal} {J. Phys. Soc. Jpn.}\ }\textbf {\bibinfo {volume}
  {84}},\ \bibinfo {pages} {014707} (\bibinfo {year} {2014})}\BibitemShut
  {NoStop}%
\bibitem [{\citenamefont {Ying}\ \emph {et~al.}(2013)\citenamefont {Ying},
  \citenamefont {Staley}, \citenamefont {Xin}, \citenamefont {Sun},
  \citenamefont {Cai}, \citenamefont {Fobes}, \citenamefont {Liu},
  \citenamefont {Mao},\ and\ \citenamefont {Liu}}]{Ying}%
  \BibitemOpen
  \bibfield  {author} {\bibinfo {author} {\bibfnamefont {Y.~A.}\ \bibnamefont
  {Ying}}, \bibinfo {author} {\bibfnamefont {N.~E.}\ \bibnamefont {Staley}},
  \bibinfo {author} {\bibfnamefont {Y.}~\bibnamefont {Xin}}, \bibinfo {author}
  {\bibfnamefont {K.}~\bibnamefont {Sun}}, \bibinfo {author} {\bibfnamefont
  {X.}~\bibnamefont {Cai}}, \bibinfo {author} {\bibfnamefont {D.}~\bibnamefont
  {Fobes}}, \bibinfo {author} {\bibfnamefont {T.~J.}\ \bibnamefont {Liu}},
  \bibinfo {author} {\bibfnamefont {Z.~Q.}\ \bibnamefont {Mao}}, \ and\
  \bibinfo {author} {\bibfnamefont {Y.}~\bibnamefont {Liu}},\ }\href
  {http://dx.doi.org/10.1038/ncomms3596} {\bibfield  {journal} {\bibinfo
  {journal} {Nat. Commun.}\ }\textbf {\bibinfo {volume} {4}},\  (\bibinfo
  {year} {2013})}\BibitemShut {NoStop}%
\bibitem [{\citenamefont {Hicks}\ \emph {et~al.}(2014)\citenamefont {Hicks},
  \citenamefont {Brodsky}, \citenamefont {Yelland}, \citenamefont {Gibbs},
  \citenamefont {Bruin}, \citenamefont {Barber}, \citenamefont {Edkins},
  \citenamefont {Nishimura}, \citenamefont {Yonezawa}, \citenamefont {Maeno},\
  and\ \citenamefont {Mackenzie}}]{Hicks}%
  \BibitemOpen
  \bibfield  {author} {\bibinfo {author} {\bibfnamefont {C.~W.}\ \bibnamefont
  {Hicks}}, \bibinfo {author} {\bibfnamefont {D.~O.}\ \bibnamefont {Brodsky}},
  \bibinfo {author} {\bibfnamefont {E.~A.}\ \bibnamefont {Yelland}}, \bibinfo
  {author} {\bibfnamefont {A.~S.}\ \bibnamefont {Gibbs}}, \bibinfo {author}
  {\bibfnamefont {J.~A.~N.}\ \bibnamefont {Bruin}}, \bibinfo {author}
  {\bibfnamefont {M.~E.}\ \bibnamefont {Barber}}, \bibinfo {author}
  {\bibfnamefont {S.~D.}\ \bibnamefont {Edkins}}, \bibinfo {author}
  {\bibfnamefont {K.}~\bibnamefont {Nishimura}}, \bibinfo {author}
  {\bibfnamefont {S.}~\bibnamefont {Yonezawa}}, \bibinfo {author}
  {\bibfnamefont {Y.}~\bibnamefont {Maeno}}, \ and\ \bibinfo {author}
  {\bibfnamefont {A.~P.}\ \bibnamefont {Mackenzie}},\ }\href {\doibase
  10.1126/science.1248292} {\bibfield  {journal} {\bibinfo  {journal}
  {Science}\ }\textbf {\bibinfo {volume} {344}},\ \bibinfo {pages} {283}
  (\bibinfo {year} {2014})}\BibitemShut {NoStop}%
\bibitem [{\citenamefont {Steppke}\ \emph {et~al.}(2016)\citenamefont
  {Steppke}, \citenamefont {Zhao}, \citenamefont {Barber}, \citenamefont
  {Scaffidi}, \citenamefont {Jerzembeck}, \citenamefont {Rosner}, \citenamefont
  {Gibbs}, \citenamefont {Maeno}, \citenamefont {Simon}, \citenamefont
  {Mackenzie},\ and\ \citenamefont {Hicks}}]{Steppke}%
  \BibitemOpen
  \bibfield  {author} {\bibinfo {author} {\bibfnamefont {A.}~\bibnamefont
  {Steppke}}, \bibinfo {author} {\bibfnamefont {L.}~\bibnamefont {Zhao}},
  \bibinfo {author} {\bibfnamefont {M.~E.}\ \bibnamefont {Barber}}, \bibinfo
  {author} {\bibfnamefont {T.}~\bibnamefont {Scaffidi}}, \bibinfo {author}
  {\bibfnamefont {F.}~\bibnamefont {Jerzembeck}}, \bibinfo {author}
  {\bibfnamefont {H.}~\bibnamefont {Rosner}}, \bibinfo {author} {\bibfnamefont
  {A.~S.}\ \bibnamefont {Gibbs}}, \bibinfo {author} {\bibfnamefont
  {Y.}~\bibnamefont {Maeno}}, \bibinfo {author} {\bibfnamefont {S.~H.}\
  \bibnamefont {Simon}}, \bibinfo {author} {\bibfnamefont {A.~P.}\ \bibnamefont
  {Mackenzie}}, \ and\ \bibinfo {author} {\bibfnamefont {C.~W.}\ \bibnamefont
  {Hicks}},\ }\href@noop {} {\  (\bibinfo {year} {2016})},\ \Eprint
  {http://arxiv.org/abs/arXiv:1604.06669} {arXiv:1604.06669} \BibitemShut
  {NoStop}%
\bibitem [{\citenamefont {Liu}\ \emph {et~al.}(2016)\citenamefont {Liu},
  \citenamefont {Zhang}, \citenamefont {Rice},\ and\ \citenamefont
  {Wang}}]{Liu}%
  \BibitemOpen
  \bibfield  {author} {\bibinfo {author} {\bibfnamefont {Y.~C.}\ \bibnamefont
  {Liu}}, \bibinfo {author} {\bibfnamefont {F.~C.}\ \bibnamefont {Zhang}},
  \bibinfo {author} {\bibfnamefont {T.~M.}\ \bibnamefont {Rice}}, \ and\
  \bibinfo {author} {\bibfnamefont {Q.~H.}\ \bibnamefont {Wang}},\ }\href@noop
  {} {\  (\bibinfo {year} {2016})},\ \Eprint
  {http://arxiv.org/abs/arXiv:1604.06666} {arXiv:1604.06666} \BibitemShut
  {NoStop}%
\bibitem [{\citenamefont {Mackenzie}\ \emph {et~al.}(1998)\citenamefont
  {Mackenzie}, \citenamefont {Haselwimmer}, \citenamefont {Tyler},
  \citenamefont {Lonzarich}, \citenamefont {Mori}, \citenamefont {Nishizaki},\
  and\ \citenamefont {Maeno}}]{Haselwimmer}%
  \BibitemOpen
  \bibfield  {author} {\bibinfo {author} {\bibfnamefont {A.~P.}\ \bibnamefont
  {Mackenzie}}, \bibinfo {author} {\bibfnamefont {R.~K.~W.}\ \bibnamefont
  {Haselwimmer}}, \bibinfo {author} {\bibfnamefont {A.~W.}\ \bibnamefont
  {Tyler}}, \bibinfo {author} {\bibfnamefont {G.~G.}\ \bibnamefont
  {Lonzarich}}, \bibinfo {author} {\bibfnamefont {Y.}~\bibnamefont {Mori}},
  \bibinfo {author} {\bibfnamefont {S.}~\bibnamefont {Nishizaki}}, \ and\
  \bibinfo {author} {\bibfnamefont {Y.}~\bibnamefont {Maeno}},\ }\href
  {\doibase 10.1103/PhysRevLett.80.161} {\bibfield  {journal} {\bibinfo
  {journal} {Phys. Rev. Lett.}\ }\textbf {\bibinfo {volume} {80}},\ \bibinfo
  {pages} {161} (\bibinfo {year} {1998})}\BibitemShut {NoStop}%
\bibitem [{\citenamefont {Agterberg}\ \emph {et~al.}(1997)\citenamefont
  {Agterberg}, \citenamefont {Rice},\ and\ \citenamefont
  {Sigrist}}]{Agterberg}%
  \BibitemOpen
  \bibfield  {author} {\bibinfo {author} {\bibfnamefont {D.~F.}\ \bibnamefont
  {Agterberg}}, \bibinfo {author} {\bibfnamefont {T.~M.}\ \bibnamefont {Rice}},
  \ and\ \bibinfo {author} {\bibfnamefont {M.}~\bibnamefont {Sigrist}},\ }\href
  {\doibase 10.1103/PhysRevLett.78.3374} {\bibfield  {journal} {\bibinfo
  {journal} {Phys. Rev. Lett.}\ }\textbf {\bibinfo {volume} {78}},\ \bibinfo
  {pages} {3374} (\bibinfo {year} {1997})}\BibitemShut {NoStop}%
\bibitem [{\citenamefont {Scaffidi}\ \emph
  {et~al.}(2014{\natexlab{a}})\citenamefont {Scaffidi}, \citenamefont
  {Romers},\ and\ \citenamefont {Simon}}]{Scaffidi}%
  \BibitemOpen
  \bibfield  {author} {\bibinfo {author} {\bibfnamefont {T.}~\bibnamefont
  {Scaffidi}}, \bibinfo {author} {\bibfnamefont {J.~C.}\ \bibnamefont
  {Romers}}, \ and\ \bibinfo {author} {\bibfnamefont {S.~H.}\ \bibnamefont
  {Simon}},\ }\href {\doibase 10.1103/PhysRevB.89.220510} {\bibfield  {journal}
  {\bibinfo  {journal} {Phys. Rev. B}\ }\textbf {\bibinfo {volume} {89}},\
  \bibinfo {pages} {220510} (\bibinfo {year} {2014}{\natexlab{a}})}\BibitemShut
  {NoStop}%
\bibitem [{\citenamefont {Firmo}\ \emph {et~al.}(2013)\citenamefont {Firmo},
  \citenamefont {Lederer}, \citenamefont {Lupien}, \citenamefont {Mackenzie},
  \citenamefont {Davis},\ and\ \citenamefont {Kivelson}}]{Firmo:2013aa}%
  \BibitemOpen
  \bibfield  {author} {\bibinfo {author} {\bibfnamefont {I.~A.}\ \bibnamefont
  {Firmo}}, \bibinfo {author} {\bibfnamefont {S.}~\bibnamefont {Lederer}},
  \bibinfo {author} {\bibfnamefont {C.}~\bibnamefont {Lupien}}, \bibinfo
  {author} {\bibfnamefont {A.~P.}\ \bibnamefont {Mackenzie}}, \bibinfo {author}
  {\bibfnamefont {J.~C.}\ \bibnamefont {Davis}}, \ and\ \bibinfo {author}
  {\bibfnamefont {S.~A.}\ \bibnamefont {Kivelson}},\ }\href {\doibase
  10.1103/PhysRevB.88.134521} {\bibfield  {journal} {\bibinfo  {journal} {Phys.
  Rev. B}\ }\textbf {\bibinfo {volume} {88}},\ \bibinfo {pages} {134521}
  (\bibinfo {year} {2013})}\BibitemShut {NoStop}%
\bibitem [{\citenamefont {Nomura}\ and\ \citenamefont
  {Yamada}(2000)}]{Nomura:2000aa}%
  \BibitemOpen
  \bibfield  {author} {\bibinfo {author} {\bibfnamefont {T.}~\bibnamefont
  {Nomura}}\ and\ \bibinfo {author} {\bibfnamefont {K.}~\bibnamefont
  {Yamada}},\ }\href {\doibase 10.1143/JPSJ.69.3678} {\bibfield  {journal}
  {\bibinfo  {journal} {Journal of the Physical Society of Japan}\ }\textbf
  {\bibinfo {volume} {69}},\ \bibinfo {pages} {3678} (\bibinfo {year}
  {2000})}\BibitemShut {NoStop}%
\bibitem [{\citenamefont {Wang}\ \emph
  {et~al.}(2013{\natexlab{b}})\citenamefont {Wang}, \citenamefont {Platt},
  \citenamefont {Yang}, \citenamefont {Honerkamp}, \citenamefont {Zhang},
  \citenamefont {Hanke}, \citenamefont {Rice},\ and\ \citenamefont
  {Thomale}}]{Wang:2013aa}%
  \BibitemOpen
  \bibfield  {author} {\bibinfo {author} {\bibfnamefont {Q.~H.}\ \bibnamefont
  {Wang}}, \bibinfo {author} {\bibfnamefont {C.}~\bibnamefont {Platt}},
  \bibinfo {author} {\bibfnamefont {Y.}~\bibnamefont {Yang}}, \bibinfo {author}
  {\bibfnamefont {C.}~\bibnamefont {Honerkamp}}, \bibinfo {author}
  {\bibfnamefont {F.~C.}\ \bibnamefont {Zhang}}, \bibinfo {author}
  {\bibfnamefont {W.}~\bibnamefont {Hanke}}, \bibinfo {author} {\bibfnamefont
  {T.~M.}\ \bibnamefont {Rice}}, \ and\ \bibinfo {author} {\bibfnamefont
  {R.}~\bibnamefont {Thomale}},\ }\href
  {http://stacks.iop.org/0295-5075/104/i=1/a=17013} {\bibfield  {journal}
  {\bibinfo  {journal} {EPL (Europhysics Letters)}\ }\textbf {\bibinfo {volume}
  {104}},\ \bibinfo {pages} {17013} (\bibinfo {year}
  {2013}{\natexlab{b}})}\BibitemShut {NoStop}%
\bibitem [{\citenamefont {Yanase}\ and\ \citenamefont
  {Ogata}(2003)}]{Yanase:2003aa}%
  \BibitemOpen
  \bibfield  {author} {\bibinfo {author} {\bibfnamefont {Y.}~\bibnamefont
  {Yanase}}\ and\ \bibinfo {author} {\bibfnamefont {M.}~\bibnamefont {Ogata}},\
  }\href {\doibase 10.1143/JPSJ.72.673} {\bibfield  {journal} {\bibinfo
  {journal} {Journal of the Physical Society of Japan}\ }\textbf {\bibinfo
  {volume} {72}},\ \bibinfo {pages} {673} (\bibinfo {year} {2003})}\BibitemShut
  {NoStop}%
\bibitem [{\citenamefont {Nomura}\ and\ \citenamefont
  {Yamada}(2002)}]{Nomura:2002aa}%
  \BibitemOpen
  \bibfield  {author} {\bibinfo {author} {\bibfnamefont {T.}~\bibnamefont
  {Nomura}}\ and\ \bibinfo {author} {\bibfnamefont {K.}~\bibnamefont
  {Yamada}},\ }\href {\doibase 10.1143/JPSJ.71.404} {\bibfield  {journal}
  {\bibinfo  {journal} {Journal of the Physical Society of Japan}\ }\textbf
  {\bibinfo {volume} {71}},\ \bibinfo {pages} {404} (\bibinfo {year}
  {2002})}\BibitemShut {NoStop}%
\bibitem [{\citenamefont {Mineev}(2014)}]{Mineev}%
  \BibitemOpen
  \bibfield  {author} {\bibinfo {author} {\bibfnamefont {V.}~\bibnamefont
  {Mineev}},\ }\href@noop {} {\bibfield  {journal} {\bibinfo  {journal} {Phys.
  Rev. B}\ }\textbf {\bibinfo {volume} {89}},\ \bibinfo {pages} {134519}
  (\bibinfo {year} {2014})}\BibitemShut {NoStop}%
\bibitem [{\citenamefont {Braden}\ \emph {et~al.}(2004)\citenamefont {Braden},
  \citenamefont {Steffens}, \citenamefont {Sidis}, \citenamefont {Kulda},
  \citenamefont {Bourges}, \citenamefont {Hayden}, \citenamefont {Kikugawa},\
  and\ \citenamefont {Maeno}}]{Sidis}%
  \BibitemOpen
  \bibfield  {author} {\bibinfo {author} {\bibfnamefont {M.}~\bibnamefont
  {Braden}}, \bibinfo {author} {\bibfnamefont {P.}~\bibnamefont {Steffens}},
  \bibinfo {author} {\bibfnamefont {Y.}~\bibnamefont {Sidis}}, \bibinfo
  {author} {\bibfnamefont {J.}~\bibnamefont {Kulda}}, \bibinfo {author}
  {\bibfnamefont {P.}~\bibnamefont {Bourges}}, \bibinfo {author} {\bibfnamefont
  {S.}~\bibnamefont {Hayden}}, \bibinfo {author} {\bibfnamefont
  {N.}~\bibnamefont {Kikugawa}}, \ and\ \bibinfo {author} {\bibfnamefont
  {Y.}~\bibnamefont {Maeno}},\ }\href@noop {} {\bibfield  {journal} {\bibinfo
  {journal} {Phys. Rev. Lett.}\ }\textbf {\bibinfo {volume} {92}},\ \bibinfo
  {pages} {097402} (\bibinfo {year} {2004})}\BibitemShut {NoStop}%
\bibitem [{\citenamefont {Veenstra}\ \emph {et~al.}(2014)\citenamefont
  {Veenstra}, \citenamefont {Zhu}, \citenamefont {Raichle}, \citenamefont
  {Ludbrook}, \citenamefont {Nicolaou}, \citenamefont {Slomski}, \citenamefont
  {Landolt}, \citenamefont {Kittaka}, \citenamefont {Maeno}, \citenamefont
  {Dil}, \citenamefont {Elfimov}, \citenamefont {Haverkort},\ and\
  \citenamefont {Damascelli}}]{Veenstra2014}%
  \BibitemOpen
  \bibfield  {author} {\bibinfo {author} {\bibfnamefont {C.~N.}\ \bibnamefont
  {Veenstra}}, \bibinfo {author} {\bibfnamefont {Z.-H.}\ \bibnamefont {Zhu}},
  \bibinfo {author} {\bibfnamefont {M.}~\bibnamefont {Raichle}}, \bibinfo
  {author} {\bibfnamefont {B.~M.}\ \bibnamefont {Ludbrook}}, \bibinfo {author}
  {\bibfnamefont {A.}~\bibnamefont {Nicolaou}}, \bibinfo {author}
  {\bibfnamefont {B.}~\bibnamefont {Slomski}}, \bibinfo {author} {\bibfnamefont
  {G.}~\bibnamefont {Landolt}}, \bibinfo {author} {\bibfnamefont
  {S.}~\bibnamefont {Kittaka}}, \bibinfo {author} {\bibfnamefont
  {Y.}~\bibnamefont {Maeno}}, \bibinfo {author} {\bibfnamefont {J.~H.}\
  \bibnamefont {Dil}}, \bibinfo {author} {\bibfnamefont {I.~S.}\ \bibnamefont
  {Elfimov}}, \bibinfo {author} {\bibfnamefont {M.~W.}\ \bibnamefont
  {Haverkort}}, \ and\ \bibinfo {author} {\bibfnamefont {A.}~\bibnamefont
  {Damascelli}},\ }\href {\doibase 10.1103/PhysRevLett.112.127002} {\bibfield
  {journal} {\bibinfo  {journal} {Phys. Rev. Lett.}\ }\textbf {\bibinfo
  {volume} {112}},\ \bibinfo {pages} {127002} (\bibinfo {year}
  {2014})}\BibitemShut {NoStop}%
\bibitem [{\citenamefont {Ng}\ and\ \citenamefont {Sigrist}(2000)}]{Ng}%
  \BibitemOpen
  \bibfield  {author} {\bibinfo {author} {\bibfnamefont {K.~K.}\ \bibnamefont
  {Ng}}\ and\ \bibinfo {author} {\bibfnamefont {M.}~\bibnamefont {Sigrist}},\
  }\href {http://stacks.iop.org/0295-5075/49/i=4/a=473} {\bibfield  {journal}
  {\bibinfo  {journal} {EPL (Europhysics Letters)}\ }\textbf {\bibinfo {volume}
  {49}},\ \bibinfo {pages} {473} (\bibinfo {year} {2000})}\BibitemShut
  {NoStop}%
\bibitem [{\citenamefont {Eremin}\ \emph {et~al.}(2002)\citenamefont {Eremin},
  \citenamefont {Manske},\ and\ \citenamefont {Bennemann}}]{Eremin2}%
  \BibitemOpen
  \bibfield  {author} {\bibinfo {author} {\bibfnamefont {I.}~\bibnamefont
  {Eremin}}, \bibinfo {author} {\bibfnamefont {D.}~\bibnamefont {Manske}}, \
  and\ \bibinfo {author} {\bibfnamefont {K.~H.}\ \bibnamefont {Bennemann}},\
  }\href {\doibase 10.1103/PhysRevB.65.220502} {\bibfield  {journal} {\bibinfo
  {journal} {Phys. Rev. B}\ }\textbf {\bibinfo {volume} {65}},\ \bibinfo
  {pages} {220502} (\bibinfo {year} {2002})}\BibitemShut {NoStop}%
\bibitem [{\citenamefont {Zabolotnyy}\ \emph {et~al.}(2013)\citenamefont
  {Zabolotnyy}, \citenamefont {Evtushinsky}, \citenamefont {Kordyuk},
  \citenamefont {Kim}, \citenamefont {Carleschi}, \citenamefont {Doyle},
  \citenamefont {Fittipaldi}, \citenamefont {Cuoco}, \citenamefont
  {Vecchione},\ and\ \citenamefont {Borisenko}}]{Zabolotnyy}%
  \BibitemOpen
  \bibfield  {author} {\bibinfo {author} {\bibfnamefont {V.~B.}\ \bibnamefont
  {Zabolotnyy}}, \bibinfo {author} {\bibfnamefont {D.~V.}\ \bibnamefont
  {Evtushinsky}}, \bibinfo {author} {\bibfnamefont {A.~A.}\ \bibnamefont
  {Kordyuk}}, \bibinfo {author} {\bibfnamefont {T.~K.}\ \bibnamefont {Kim}},
  \bibinfo {author} {\bibfnamefont {E.}~\bibnamefont {Carleschi}}, \bibinfo
  {author} {\bibfnamefont {B.~P.}\ \bibnamefont {Doyle}}, \bibinfo {author}
  {\bibfnamefont {R.}~\bibnamefont {Fittipaldi}}, \bibinfo {author}
  {\bibfnamefont {M.}~\bibnamefont {Cuoco}}, \bibinfo {author} {\bibfnamefont
  {A.}~\bibnamefont {Vecchione}}, \ and\ \bibinfo {author} {\bibfnamefont
  {S.~V.}\ \bibnamefont {Borisenko}},\ }\href@noop {} {\bibfield  {journal}
  {\bibinfo  {journal} {Journal of Electron Spectroscopy and Related
  Phenomena}\ }\textbf {\bibinfo {volume} {191}},\ \bibinfo {pages} {48}
  (\bibinfo {year} {2013})}\BibitemShut {NoStop}%
\bibitem [{\citenamefont {de' Medici}\ \emph {et~al.}(2011)\citenamefont {de'
  Medici}, \citenamefont {Mravlje},\ and\ \citenamefont {Georges}}]{Medici}%
  \BibitemOpen
  \bibfield  {author} {\bibinfo {author} {\bibfnamefont {L.}~\bibnamefont {de'
  Medici}}, \bibinfo {author} {\bibfnamefont {J.}~\bibnamefont {Mravlje}}, \
  and\ \bibinfo {author} {\bibfnamefont {A.}~\bibnamefont {Georges}},\ }\href
  {\doibase 10.1103/PhysRevLett.107.256401} {\bibfield  {journal} {\bibinfo
  {journal} {Phys. Rev. Lett.}\ }\textbf {\bibinfo {volume} {107}},\ \bibinfo
  {pages} {256401} (\bibinfo {year} {2011})}\BibitemShut {NoStop}%
\bibitem [{\citenamefont {Mravlje}\ \emph {et~al.}(2011)\citenamefont
  {Mravlje}, \citenamefont {Aichhorn}, \citenamefont {Miyake}, \citenamefont
  {Haule}, \citenamefont {Kotliar},\ and\ \citenamefont {Georges}}]{Aichhorn}%
  \BibitemOpen
  \bibfield  {author} {\bibinfo {author} {\bibfnamefont {J.}~\bibnamefont
  {Mravlje}}, \bibinfo {author} {\bibfnamefont {M.}~\bibnamefont {Aichhorn}},
  \bibinfo {author} {\bibfnamefont {T.}~\bibnamefont {Miyake}}, \bibinfo
  {author} {\bibfnamefont {K.}~\bibnamefont {Haule}}, \bibinfo {author}
  {\bibfnamefont {G.}~\bibnamefont {Kotliar}}, \ and\ \bibinfo {author}
  {\bibfnamefont {A.}~\bibnamefont {Georges}},\ }\href {\doibase
  10.1103/PhysRevLett.106.096401} {\bibfield  {journal} {\bibinfo  {journal}
  {Phys. Rev. Lett.}\ }\textbf {\bibinfo {volume} {106}},\ \bibinfo {pages}
  {096401} (\bibinfo {year} {2011})}\BibitemShut {NoStop}%
\bibitem [{\citenamefont {Georges}\ \emph {et~al.}(2013)\citenamefont
  {Georges}, \citenamefont {de' Medici},\ and\ \citenamefont
  {Mravlje}}]{Medici2}%
  \BibitemOpen
  \bibfield  {author} {\bibinfo {author} {\bibfnamefont {A.}~\bibnamefont
  {Georges}}, \bibinfo {author} {\bibfnamefont {L.}~\bibnamefont {de' Medici}},
  \ and\ \bibinfo {author} {\bibfnamefont {J.}~\bibnamefont {Mravlje}},\ }\href
  {\doibase 10.1146/annurev-conmatphys-020911-125045} {\bibfield  {journal}
  {\bibinfo  {journal} {Annual Review of Condensed Matter Physics}\ }\textbf
  {\bibinfo {volume} {4}},\ \bibinfo {pages} {137} (\bibinfo {year}
  {2013})}\BibitemShut {NoStop}%
\bibitem [{\citenamefont {Ryee}\ \emph {et~al.}(2016)\citenamefont {Ryee},
  \citenamefont {Jang}, \citenamefont {Kino}, \citenamefont {Kotani},\ and\
  \citenamefont {Han}}]{Ryee}%
  \BibitemOpen
  \bibfield  {author} {\bibinfo {author} {\bibfnamefont {S.}~\bibnamefont
  {Ryee}}, \bibinfo {author} {\bibfnamefont {S.~W.}\ \bibnamefont {Jang}},
  \bibinfo {author} {\bibfnamefont {H.}~\bibnamefont {Kino}}, \bibinfo {author}
  {\bibfnamefont {T.}~\bibnamefont {Kotani}}, \ and\ \bibinfo {author}
  {\bibfnamefont {M.~J.}\ \bibnamefont {Han}},\ }\href {\doibase
  10.1103/PhysRevB.93.075125} {\bibfield  {journal} {\bibinfo  {journal} {Phys.
  Rev. B}\ }\textbf {\bibinfo {volume} {93}},\ \bibinfo {pages} {075125}
  (\bibinfo {year} {2016})}\BibitemShut {NoStop}%
\bibitem [{\citenamefont {Tran}\ \emph {et~al.}(2004)\citenamefont {Tran},
  \citenamefont {Mizokawa}, \citenamefont {Nakatsuji}, \citenamefont
  {Fukazawa},\ and\ \citenamefont {Maeno}}]{Tran}%
  \BibitemOpen
  \bibfield  {author} {\bibinfo {author} {\bibfnamefont {T.~T.}\ \bibnamefont
  {Tran}}, \bibinfo {author} {\bibfnamefont {T.}~\bibnamefont {Mizokawa}},
  \bibinfo {author} {\bibfnamefont {S.}~\bibnamefont {Nakatsuji}}, \bibinfo
  {author} {\bibfnamefont {H.}~\bibnamefont {Fukazawa}}, \ and\ \bibinfo
  {author} {\bibfnamefont {Y.}~\bibnamefont {Maeno}},\ }\href {\doibase
  10.1103/PhysRevB.70.153106} {\bibfield  {journal} {\bibinfo  {journal} {Phys.
  Rev. B}\ }\textbf {\bibinfo {volume} {70}},\ \bibinfo {pages} {153106}
  (\bibinfo {year} {2004})}\BibitemShut {NoStop}%
\bibitem [{\citenamefont {Pchelkina}\ \emph {et~al.}(2007)\citenamefont
  {Pchelkina}, \citenamefont {Nekrasov}, \citenamefont {Pruschke},
  \citenamefont {Sekiyama}, \citenamefont {Suga}, \citenamefont {Anisimov},\
  and\ \citenamefont {Vollhardt}}]{Pchelkina}%
  \BibitemOpen
  \bibfield  {author} {\bibinfo {author} {\bibfnamefont {Z.~V.}\ \bibnamefont
  {Pchelkina}}, \bibinfo {author} {\bibfnamefont {I.~A.}\ \bibnamefont
  {Nekrasov}}, \bibinfo {author} {\bibfnamefont {T.}~\bibnamefont {Pruschke}},
  \bibinfo {author} {\bibfnamefont {A.}~\bibnamefont {Sekiyama}}, \bibinfo
  {author} {\bibfnamefont {S.}~\bibnamefont {Suga}}, \bibinfo {author}
  {\bibfnamefont {V.~I.}\ \bibnamefont {Anisimov}}, \ and\ \bibinfo {author}
  {\bibfnamefont {D.}~\bibnamefont {Vollhardt}},\ }\href {\doibase
  10.1103/PhysRevB.75.035122} {\bibfield  {journal} {\bibinfo  {journal} {Phys.
  Rev. B}\ }\textbf {\bibinfo {volume} {75}},\ \bibinfo {pages} {035122}
  (\bibinfo {year} {2007})}\BibitemShut {NoStop}%
\bibitem [{\citenamefont {Singh}(2008)}]{Singh}%
  \BibitemOpen
  \bibfield  {author} {\bibinfo {author} {\bibfnamefont {D.~J.}\ \bibnamefont
  {Singh}},\ }\href {\doibase 10.1103/PhysRevB.77.046101} {\bibfield  {journal}
  {\bibinfo  {journal} {Phys. Rev. B}\ }\textbf {\bibinfo {volume} {77}},\
  \bibinfo {pages} {046101} (\bibinfo {year} {2008})}\BibitemShut {NoStop}%
\bibitem [{\citenamefont {Pchelkina}\ \emph {et~al.}(2008)\citenamefont
  {Pchelkina}, \citenamefont {Nekrasov}, \citenamefont {Pruschke},
  \citenamefont {Suga}, \citenamefont {Anisimov},\ and\ \citenamefont
  {Vollhardt}}]{Pchelkina2}%
  \BibitemOpen
  \bibfield  {author} {\bibinfo {author} {\bibfnamefont {Z.~V.}\ \bibnamefont
  {Pchelkina}}, \bibinfo {author} {\bibfnamefont {I.~A.}\ \bibnamefont
  {Nekrasov}}, \bibinfo {author} {\bibfnamefont {T.}~\bibnamefont {Pruschke}},
  \bibinfo {author} {\bibfnamefont {S.}~\bibnamefont {Suga}}, \bibinfo {author}
  {\bibfnamefont {V.~I.}\ \bibnamefont {Anisimov}}, \ and\ \bibinfo {author}
  {\bibfnamefont {D.}~\bibnamefont {Vollhardt}},\ }\href {\doibase
  10.1103/PhysRevB.77.046102} {\bibfield  {journal} {\bibinfo  {journal} {Phys.
  Rev. B}\ }\textbf {\bibinfo {volume} {77}},\ \bibinfo {pages} {046102}
  (\bibinfo {year} {2008})}\BibitemShut {NoStop}%
\bibitem [{\citenamefont {Damascelli}\ \emph {et~al.}(2000)\citenamefont
  {Damascelli}, \citenamefont {Lu}, \citenamefont {Shen}, \citenamefont
  {Armitage}, \citenamefont {Ronning}, \citenamefont {Feng}, \citenamefont
  {Kim}, \citenamefont {Shen}, \citenamefont {Kimura}, \citenamefont {Tokura},
  \citenamefont {Mao},\ and\ \citenamefont {Maeno}}]{Damascelli}%
  \BibitemOpen
  \bibfield  {author} {\bibinfo {author} {\bibfnamefont {A.}~\bibnamefont
  {Damascelli}}, \bibinfo {author} {\bibfnamefont {D.~H.}\ \bibnamefont {Lu}},
  \bibinfo {author} {\bibfnamefont {K.~M.}\ \bibnamefont {Shen}}, \bibinfo
  {author} {\bibfnamefont {N.~P.}\ \bibnamefont {Armitage}}, \bibinfo {author}
  {\bibfnamefont {F.}~\bibnamefont {Ronning}}, \bibinfo {author} {\bibfnamefont
  {D.~L.}\ \bibnamefont {Feng}}, \bibinfo {author} {\bibfnamefont
  {C.}~\bibnamefont {Kim}}, \bibinfo {author} {\bibfnamefont {Z.-X.}\
  \bibnamefont {Shen}}, \bibinfo {author} {\bibfnamefont {T.}~\bibnamefont
  {Kimura}}, \bibinfo {author} {\bibfnamefont {Y.}~\bibnamefont {Tokura}},
  \bibinfo {author} {\bibfnamefont {Z.~Q.}\ \bibnamefont {Mao}}, \ and\
  \bibinfo {author} {\bibfnamefont {Y.}~\bibnamefont {Maeno}},\ }\href
  {\doibase 10.1103/PhysRevLett.85.5194} {\bibfield  {journal} {\bibinfo
  {journal} {Phys. Rev. Lett.}\ }\textbf {\bibinfo {volume} {85}},\ \bibinfo
  {pages} {5194} (\bibinfo {year} {2000})}\BibitemShut {NoStop}%
\bibitem [{\citenamefont {Mackenzie}\ \emph {et~al.}(1996)\citenamefont
  {Mackenzie}, \citenamefont {Julian}, \citenamefont {Diver}, \citenamefont
  {McMullan}, \citenamefont {Ray}, \citenamefont {Lonzarich}, \citenamefont
  {Maeno}, \citenamefont {Nishizaki},\ and\ \citenamefont
  {Fujita}}]{Mackenzie3}%
  \BibitemOpen
  \bibfield  {author} {\bibinfo {author} {\bibfnamefont {A.~P.}\ \bibnamefont
  {Mackenzie}}, \bibinfo {author} {\bibfnamefont {S.~R.}\ \bibnamefont
  {Julian}}, \bibinfo {author} {\bibfnamefont {A.~J.}\ \bibnamefont {Diver}},
  \bibinfo {author} {\bibfnamefont {G.~J.}\ \bibnamefont {McMullan}}, \bibinfo
  {author} {\bibfnamefont {M.~P.}\ \bibnamefont {Ray}}, \bibinfo {author}
  {\bibfnamefont {G.~G.}\ \bibnamefont {Lonzarich}}, \bibinfo {author}
  {\bibfnamefont {Y.}~\bibnamefont {Maeno}}, \bibinfo {author} {\bibfnamefont
  {S.}~\bibnamefont {Nishizaki}}, \ and\ \bibinfo {author} {\bibfnamefont
  {T.}~\bibnamefont {Fujita}},\ }\href {\doibase 10.1103/PhysRevLett.76.3786}
  {\bibfield  {journal} {\bibinfo  {journal} {Phys. Rev. Lett.}\ }\textbf
  {\bibinfo {volume} {76}},\ \bibinfo {pages} {3786} (\bibinfo {year}
  {1996})}\BibitemShut {NoStop}%
\bibitem [{\citenamefont {Bergemann}\ \emph {et~al.}(2003)\citenamefont
  {Bergemann}, \citenamefont {Mackenzie}, \citenamefont {Julian}, \citenamefont
  {Forsythe},\ and\ \citenamefont {Ohmichi}}]{Bergemann}%
  \BibitemOpen
  \bibfield  {author} {\bibinfo {author} {\bibfnamefont {C.}~\bibnamefont
  {Bergemann}}, \bibinfo {author} {\bibfnamefont {A.~P.}\ \bibnamefont
  {Mackenzie}}, \bibinfo {author} {\bibfnamefont {S.~R.}\ \bibnamefont
  {Julian}}, \bibinfo {author} {\bibfnamefont {D.}~\bibnamefont {Forsythe}}, \
  and\ \bibinfo {author} {\bibfnamefont {E.}~\bibnamefont {Ohmichi}},\
  }\bibfield  {booktitle} {\emph {\bibinfo {booktitle} {Advances in Physics}},\
  }\href {\doibase 10.1080/00018730310001621737} {\bibfield  {journal}
  {\bibinfo  {journal} {Advances in Physics}\ }\textbf {\bibinfo {volume}
  {52}},\ \bibinfo {pages} {639} (\bibinfo {year} {2003})}\BibitemShut
  {NoStop}%
\bibitem [{\citenamefont {Singh}(1995)}]{DavidSingh}%
  \BibitemOpen
  \bibfield  {author} {\bibinfo {author} {\bibfnamefont {D.~J.}\ \bibnamefont
  {Singh}},\ }\href {\doibase 10.1103/PhysRevB.52.1358} {\bibfield  {journal}
  {\bibinfo  {journal} {Phys. Rev. B}\ }\textbf {\bibinfo {volume} {52}},\
  \bibinfo {pages} {1358} (\bibinfo {year} {1995})}\BibitemShut {NoStop}%
\bibitem [{\citenamefont {Oguchi}(1995)}]{Oguchi}%
  \BibitemOpen
  \bibfield  {author} {\bibinfo {author} {\bibfnamefont {T.}~\bibnamefont
  {Oguchi}},\ }\href {\doibase 10.1103/PhysRevB.51.1385} {\bibfield  {journal}
  {\bibinfo  {journal} {Phys. Rev. B}\ }\textbf {\bibinfo {volume} {51}},\
  \bibinfo {pages} {1385} (\bibinfo {year} {1995})}\BibitemShut {NoStop}%
\bibitem [{\citenamefont {Scaffidi}\ \emph
  {et~al.}(2014{\natexlab{b}})\citenamefont {Scaffidi}, \citenamefont
  {Romers},\ and\ \citenamefont {Simon}}]{Simon}%
  \BibitemOpen
  \bibfield  {author} {\bibinfo {author} {\bibfnamefont {T.}~\bibnamefont
  {Scaffidi}}, \bibinfo {author} {\bibfnamefont {J.~C.}\ \bibnamefont
  {Romers}}, \ and\ \bibinfo {author} {\bibfnamefont {S.~H.}\ \bibnamefont
  {Simon}},\ }\href {\doibase 10.1103/PhysRevB.89.220510} {\bibfield  {journal}
  {\bibinfo  {journal} {Phys. Rev. B}\ }\textbf {\bibinfo {volume} {89}},\
  \bibinfo {pages} {220510} (\bibinfo {year} {2014}{\natexlab{b}})}\BibitemShut
  {NoStop}%
\bibitem [{\citenamefont {Braden}\ \emph {et~al.}(2002)\citenamefont {Braden},
  \citenamefont {Sidis}, \citenamefont {Bourges}, \citenamefont {Pfeuty},
  \citenamefont {Kulda}, \citenamefont {Mao},\ and\ \citenamefont
  {Maeno}}]{Braden2}%
  \BibitemOpen
  \bibfield  {author} {\bibinfo {author} {\bibfnamefont {M.}~\bibnamefont
  {Braden}}, \bibinfo {author} {\bibfnamefont {Y.}~\bibnamefont {Sidis}},
  \bibinfo {author} {\bibfnamefont {P.}~\bibnamefont {Bourges}}, \bibinfo
  {author} {\bibfnamefont {P.}~\bibnamefont {Pfeuty}}, \bibinfo {author}
  {\bibfnamefont {J.}~\bibnamefont {Kulda}}, \bibinfo {author} {\bibfnamefont
  {Z.}~\bibnamefont {Mao}}, \ and\ \bibinfo {author} {\bibfnamefont
  {Y.}~\bibnamefont {Maeno}},\ }\href@noop {} {\bibfield  {journal} {\bibinfo
  {journal} {Phys. Rev. B}\ }\textbf {\bibinfo {volume} {66}},\ \bibinfo
  {pages} {064522} (\bibinfo {year} {2002})}\BibitemShut {NoStop}%
\end{thebibliography}%

\end{document}